\documentclass[9pt,shortpaper,twoside,web]{ieeecolor}
\usepackage{generic}
\usepackage{cite}
\usepackage{amsmath,amssymb,amsfonts}
\usepackage{algorithmic}
\usepackage{graphicx}
\usepackage{textcomp}
\usepackage{epstopdf}
\usepackage{subfigure}
\usepackage{graphicx}
\def\BibTeX{{\rm B\kern-.05em{\sc i\kern-.025em b}\kern-.08em
    T\kern-.1667em\lower.7ex\hbox{E}\kern-.125emX}}
\begin{document}
\title{Decentralized Intermittent Feedback Adaptive Control of Non-triangular Nonlinear Time-varying Systems}
\author{Libei~Sun,~Xiucai~Huang,~Yongduan~Song,~\IEEEmembership{Fellow,~IEEE}
\thanks{The authors are with the State Key Laboratory of Power Transmission Equipment System Security and New Technology, Chongqing Key Laboratory of Intelligent Unmanned Systems, School of Automation, Chongqing University, Chongqing 400044, China (e-mail: ydsong@cqu.edu.cn, hxiucai@cqu.edu.cn, and lbsun@cqu.edu.cn). (Corresponding author: Yongduan Song).}}
\maketitle

\begin{abstract}
This paper investigates the decentralized stabilization problem for a class of interconnected systems in the presence of non-triangular structural uncertainties and time-varying parameters, where each subsystem exchanges information only with its neighbors and only intermittent (rather than continuous) states and input are to be utilized. Thus far to our best knowledge, no solution exists priori to this work, despite its high prevalence in practice. Two globally decentralized adaptive control schemes are presented based on the backstepping technique, the first one is developed in a continuous fashion by combining the philosophy of the modified congelation of variables based approach with the special treatment of non-triangular structural uncertainties, which avoids the derivative of time-varying parameters and eliminates the limitation of the triangular condition, thus largely broadens the scope of application. By making use of the important property that the partial derivatives of the constructed virtual controllers in each subsystem are all constant, the second scheme is developed through directly replacing the states in the preceding scheme with the triggered ones. Consequently, the non-differentiability of the virtual control stemming from intermittent state feedback is completely obviated. The internal signals under both schemes are rigorously shown to be globally uniformly bounded with the aid of several novel lemmas, while the stabilization performance can be enhanced by appropriately adjusting design parameters. Moreover, the inter-event intervals are ensured to be lower-bounded by a positive constant. Finally, numerical simulation verifies the benefits and efficiency of the proposed method.
\end{abstract}
\IEEEpeerreviewmaketitle
%non-triangular structural uncertainties, time-varying interconnected systems.

\section{Introduction}
%\IEEEPARstart
\allowdisplaybreaks
%Decentralized adaptive control of large-scale uncertain complex interconnected systems is currently facilitated by recent technological advances on computing and communication resources, which serves as an efficient and practical strategy to be employed for many reasons, such as simplicity of controller design and implementation \cite{wen1992global}.  ioannou1986decentralized,wen1992global,jiang2000decentralized
Large-scale uncertain complex interconnected systems are frequently encountered \cite{ioannou1986decentralized,harmand2005optimal,wen1992global,jiang2000decentralized}, decentralized adaptive control of such systems is currently facilitated by recent technological advances on computing and communication resources, which serves as an efficient and practical strategy to be employed for many reasons, such as simplicity of controller design and implementation.
Communication network is necessary for signal transmission in large-scale nonlinear control systems owing to networked control systems (NCSs) with advantages of lower cost, easier maintenance and higher reliability \cite{4118474}.
However, there is a gap between the decentralized control and the network under such framework, because the sensor data cannot be transmitted/updated in real time on account of limited communication bandwidth and channels, which potentially degrades the control performance of large-scale nonlinear system.

To preserve a trade-off between communication resource usage and control performance, the emergence of event-triggered control is stimulated as an appealing method for saving energy and communication resources, which enables communication only when certain predefined condition is triggered (see e.g., \cite{heemels2012periodic,astrom2008event} and the references therein).
Early available results on event-triggered control mainly focus on linear systems, see \cite{SEURET201647,zhu2014event} for examples. An  extension work to nonlinear systems is pioneered in \cite{tabuada2007event}, yet the closed-loop dynamic  should be input-to-state stable (ISS). Such limitation is removed in \cite{adaldo2015event} by codesigning an event-triggered control algorithm. However, the system models considered in \cite{tabuada2007event,adaldo2015event} are required to be exactly known. To handle the uncertainties for nonlinear systems, some event-triggered adaptive control schemes are advocated via the backstepping design procedure, see e.g. \cite{xing2016event,ghodrat2018local} and the references therein. Nonetheless, those results are dedicated to the case where only the control input is intermittently transmitted over the network while continuous feedback of plant states are required, thus merely saving the communication resources in the controller-to-actuator channel, but not applicable for the senor-to-controller ones.

For the past few years, control design via
intermittent state feedback has stirred an increasing amount of attention in the literature due to its more efficient usage of available limited resources. In this direction, two types of strategies should be highlighted. The first one is the state-triggered control using intermittent output only.
%One such strategy is the state-triggered output feedback control proposed in \cite{ABDELRAHIM201796} for delta operator systems with matched uncertainties.
In this direction, a state-triggered output feedback control scheme is proposed in \cite{ABDELRAHIM201796} for delta operator systems with matched uncertainties.
In \cite{ZHANG2022110283}, the problem of decentralized adaptive backstepping based output feedback control is addressed for a class of nonlinear interconnected systems. However, in both solutions, the alleviation on communication burden is still  limited because only the output is triggered. The second one is the state-triggered control via intermittent full-state feedback. For this scenario, some efforts have been made in \cite{wang2020adaptive,wang2021adaptive} by using backstepping based adaptive control, wherein the models are in low-order form  \cite{wang2020adaptive} or in normal form \cite{wang2021adaptive}. More recently, with the aid of dynamic surface control (DSC) technique, such restriction is explicitly relaxed in \cite{zhang2021adaptive} for a family of nonlinear systems with constant parametric uncertainties. The idea of designing a distributed state-triggered control algorithm for networked nonlinear system with mismatched and nonparametric uncertainties is further introduced in \cite{sun2022Distributed}.
%The multi-layered event-triggered adaptive control problem for nonlinear systems with mismatched uncertainties and unknown control gains is considered in \cite{sun2022multiply}.
The result of \cite{DEPERSIS20132116} solves the problem of stabilizing large-scale interconnected systems by distributed state-triggered controllers built on the ISS condition. Nevertheless, the approaches in \cite{zhang2021adaptive,sun2022Distributed,DEPERSIS20132116} are tailored for nonlinear systems in triangular form. Meanwhile, the plant parameters involved in the above results are all restricted to constants. In most applications, however, plant parameters may vary with time rapidly \cite{zhang2003backstepping}. For instance, traffic free speed is considered as a time-varying parameter in freeway traffic systems control, since changes in weather, air pressure, etc., can strongly influences free speed; in automatic train control problems, the mass of the train load that affects the resistance imposed on the train, may not be the same at different runs. Therefore it is vitally important to relax or even remove these strong restrictions, and broaden the applicability of the backstepping-based state-triggered stability theory to cover system models in non-triangular forms with time-varying parameters.

Motivated by the aforementioned discussion, in this work we develop a decentralized event-triggered adaptive backstepping control method for nonlinear interconnected systems with non-triangular structural uncertainties and unknown time-varying parameters. Under such setting it is actually nontrivial to achieve this goal, this is because two major technical difficulties are present in control design and stability analysis.
First, the models of subsystems are all in non-trival non-triangular form with coupling interactions and unknown time-varying parameters that directly challenges the traditional back-stepping design procedure, a tailored technique for triangular systems with constant parameters;
Second, with intermittent feedback signals arising from event-triggering, the underlying problem becomes even more complicated when carrying out backstepping design because the repetitive differentiation of virtual control signals (with respect to the triggering signals) is no longer feasible (literally undefined due to the nature of the event-triggering).
In this work, we propose two globally decentralized adaptive backstepping control design approaches respectively for the cases with and without event-triggering setting.
%
%The first one is developed in a continuous fashion that successfully removes the triangular-form-limitation imposed on the system model by refining a novel treatment on the non-triangular structural uncertainties for the backstepping design, and simultaneously restrains the affects of the parameter-induced perturbation by freezing the time-varying parameters at the centers, thus avoiding the derivation of time-varying parameters and further relaxing the related state-of-the art conditions \cite{goel2022composite,arefi2010adaptive}.
%It is shown that the partial derivatives of virtual controllers in each subsystem with respect to states are constants. With such property, the second control scheme is then constructed by replacing the states in the first one with the triggering states, thus completely overcoming the aforementioned non-differentiability problem in a global manner.
% properly treating
The first one is developed in a continuous fashion that successfully removes the triangular-form-limitation imposed on the system model by properly treating the non-triangular structural uncertainties for the backstepping design, and simultaneously restrains the affects of the parameter-induced perturbation by freezing the time-varying parameters at the centers, thus avoiding the derivative of time-varying parameters and further relaxing the related state-of-the art conditions \cite{goel2022composite,arefi2010adaptive}.
It is shown that the partial derivatives of virtual controllers in each subsystem with respect to states are constants. With such property, the second control scheme is then constructed by replacing the states in the first one with the triggering states, thus circumventing the aforementioned non-differentiability in a global manner.
Several lemmas are established to facilitate the authentication of the global uniform boundedness of all the closed-loop signals in both strategies with the stabilization performance improvable by appropriately adjusting design parameters. Moreover, a strictly positive lower bound on the inter transmission times is enforced by the proposed event triggering mechanism (ETM), thus the notorious Zeno phenomenon is avoided.
To our best knowledge, this is the first adaptive backstepping control solution for interconnected nonlinear systems under event-triggering setting that is able to tolerate non-triangular structural uncertainties and unknown time-varying parameters.

%The remainder of the paper is organized as follows. In Section II, the problem is formally stated and the needed assumptions are introduced.
%In Section III, a decentralized adaptive backstepping control scheme is firstly designed under continuous  state feedback. Based upon the preceding continuous control scheme, the decentralized event-triggered way is then proposed via intermittent state feedback in Sections IV. Section VI concludes the paper.

%${\textbf{Notation}}$.
%$\mathcal{R}$ and $\mathcal{R}^{n}$ represent the 1-$D$ real space and the $n$-dimensional real vector space, respectively, $\left\|{\cdot }\right\|$ denotes the Euclidean vector norm, $\left\|{\cdot }\right\|_F$ denotes the Frobenius norm, $\left|{\cdot }\right|$ is the absolute value of a real number, ${\rm{eig}}\left(\cdot\right)$ denotes the eigenvalue of $\left(\cdot\right)$ and $\lambda_{\max}\left(\cdot\right)$ denote the maximum eigenvalue of $\left(\cdot\right)$.

\section{Problem Formulation}%PROBLEM FORMULATION
Consider the following nonlinear system consists of $N$ interconnected subsystems, with the $i$th subsystem modeled as:
\begin{flalign}
\dot{x}_{i, k}=\,&x_{i, k+1}+\sum_{j=1}^{N} f_{i j, k}\left(x_{j}, u_{j}, t\right), k=1,\cdots, n_{i}-1&\nonumber\\
\dot{x}_{i, n_{i}}=\,&u_{i}+\varphi_{i}^{T}
\left(x_{i}\right){\theta_{i}\left(t\right)}+\psi_i\left(x_{i}\right)
+\sum_{j=1}^{N} f_{i j, n_{i}}\left(x_{j}, u_{j}, t\right)&\nonumber\\
y_{i}=\,&x_{i, 1}& \label{eq:1}
\end{flalign}
for $i=1,\cdots,N$, where ${x_{i,k}}\in \mathcal{R}$, $k = 1,\cdots ,n_i$ is the system state, with $x_i=[x_{i,1},\cdots,x_{i,n_i}]^{T}$, $u_i \in \mathcal{R}$ and $y_i\in \mathcal{R}$ are the control input and output, respectively, $\varphi_{i}\left(x_{i}\right) \in \mathcal{R}^{p}$ and $\psi_i\left(x_{i}\right) \in\mathcal{R}$ are known functions, with $\varphi_{i}\left(0\right)=0$,  ${\theta_{i}\left(t\right)}\in \mathcal{R}^{p}$ is the unknown parameter vector, $f_{i j, k}\left(x_{j}, u_{j}, t\right)\in \mathcal{R}$ denotes the nonlinear coupling interaction from the $j$th subsystem for $j \neq i$, and the modeling error of the $i$th subsystem for $j=i$  \footnote{Arguments of some functions will be omitted hereafter if no confusion is likely to occur.}.

The objective of this paper is to develop the globally decentralized adaptive backstepping control scheme for system (\ref{eq:1}) using only locally intermittent feedback signals, such that
\begin{itemize}
\item [$\bullet$]{The global uniform boundedness of the closed-loop signals is ensured, while all the subsystem outputs are steered into an assignable residual set around zero;}
\item [$\bullet$]{The Zeno behavior is precluded.}
\end{itemize}

To move on, we make the following assumptions.% with regard to dynamics (\ref{eq:1}).

${\textbf{Assumption 1}}$ \cite{cai2022decentralized}.
The unknown nonlinear function $f_{i j, k}\left(x_{j}, u_{j}, t\right)$ satisfies the following condition:
\begin{flalign}
&\left|f_{i j, k}\left(x_{j}, u_{j}, t\right)\right| \leq \hbar_{i j, k}\left\|x_{j}\right\|
+\epsilon_{i j, k}& \label{eq:2}
\end{flalign}
for $i,j=1,\cdots, N$, $k=1, \cdots, n_{i}$, where $\hbar_{ij,k}\ge{0}$ is the unknown coupling gain, which denotes the magnitudes or strengths of the modeling errors and coupling interactions, and $\epsilon_{i j, k}\ge{0}$ is an unknown constant.

${\textbf{Assumption 2}}$.
The parameter $\theta_i(t)$ is piecewise continuous and $\theta_i(t) \in \Omega_{i0}$, for all $t \geq 0$, where $\Omega_{i0}$ is an unknown compact set. The ``radius" of $\Omega_{i0}$, denoted by  $\beta_{\theta_i}$, is assumed to be bounded but not necessarily known.

${\textbf{Assumption 3}}$.
The functions $\varphi_{i}\left(x_i\right)$ and $\psi_{i}\left(x_i\right), i=1,\cdots,N$ satisfy the global Lipschitz continuity condition such that
\begin{flalign}
\left\|\varphi_{i}\left(x_i\right)
{\rm{-}}\varphi_{i}\left(\bar{x}_i\right)\right\| \leq \,& L_{\varphi_{i}}\left\|x_i-\bar{x}_i\right\| &\label{eq:df3}\\ \left|\psi_{i}\left(x_i\right)-\psi_{i}\left(\bar{x}_i\right)\right| \leq \,& L_{\psi_{i}}\left\|x_i-\bar{x}_i\right\| & \label{eq:3}
\end{flalign}
where $L_{\varphi_{i}}$ and $L_{\psi_{i}}$ are unknown bounded constants.

${\textbf{Remark 1}}$.
Notice from (\ref{eq:2}) that $f_{ij,k}\left(x_{j}, u_{j}, t\right)$ is bounded by a function that allows dependence on all subsystems states, in other words, the uncertainties under consideration fails to satisfy the triangular structure. In addition, it holds that the larger the value of $\hbar_{ij,k}$ is, the stronger the influence degree would be. Such interactions are rather general in numerous real-world systems, such as power systems, water systems, traffic systems and flexible space structures \cite{zhou2022event,pagilla2006decentralized}, which tend to degrade the system performance and thus challenge the reliability and safety of the system.

${\textbf{Remark 2}}$.
\emph{Assumption} 1 can be commonly found in literature, for example,  \cite{cai2022decentralized,cai2016adaptive}.
%Notice that the ``radius" of $\Omega_{i0}$, i.e., $\beta_{\theta_i}$ considered by \cite{chen2020adaptive} is required to be available for controller design, whereas here $\beta_{\theta_i}$ is only assumed to be bounded, as noted in \emph{Assumption} 2.Unlike the results in [12], [20]-[21]
As noted in \emph{Assumption} 2, only the ``radius" of $\Omega_{i0}$, i.e., $\beta_{\theta_i}$ is assumed to be bounded, which is more general than the existing results \cite{rios2017time,chen2020adaptive}. Specifically, it is assumed that $\dot{\theta} \in \mathcal{L}_{\infty}$ with $\|\dot{\theta}\|_{\infty} \leq h(\cdot) \leq\theta_0$ for all $t \geq 0$ in \cite{rios2017time}, where $h$ is a known continuous function and $\theta_0$ is a known positive constant. Besides, $\beta_{\theta_i}$ considered in \cite{chen2020adaptive} is required to be available for  the control design.
\emph{Assumption} 3 is quite common, see  \cite{ZHANG2022110283,zhang2021adaptive} for examples.

\section{Decentralized Continuous Adaptive Backstepping Control}
In this section, a decentralized adaptive backstepping control scheme is developed using locally continuous state signals, which can also be regarded as the basis of the control scheme via intermittent state feedback in next section. To this end, we first carry out the following change of coordinates:
\begin{flalign}
{z_{i,1}} =\,& {x_{i,1}}& \label{eq:4}\\
{z_{i,k}} =\,& {x_{i,k}} - {\alpha _{i,k-1}},k = 2, \cdots ,n_i & \label{eq:5}
\end{flalign}

The decentralized adaptive backstepping control scheme under continuous state feedback is designed as:
\begin{flalign}
\alpha_{i, 1}=\,&-c_{i, 1} z_{i, 1}-\frac{1}{4 \varpi_{i i, 1,1}} z_{i, 1}-\frac{1}{4 \varpi_{i i, 1,2}} z_{i, 1}&\nonumber \\
&-\sum_{j \neq i} \frac{1}{4 \varpi_{i j, 1,1}} z_{i, 1}-\sum_{j \neq i} \frac{1}{4 \varpi_{i j, 1,2}} z_{i, 1}& \label{eq:6}\\
\alpha_{i, k}=\,&-c_{i, k} z_{i, k}-\sum_{j=1}^{N}
\left(\frac{1}{4 \varpi_{i j, k, 1}}+\frac{1}{4 \varpi_{i j, k,2}}\right)z_{i, k}&\nonumber \\
&-\sum_{j=1}^{N}\sum_{l=1}^{k-1}
\left(\frac{\left(\xi_{k-1, l}^{i}\right)^{2}}{4 \varpi_{i j, k, 1}}+\frac{\left(\xi_{k-1, l}^{i}\right)^{2}}{4 \varpi_{i j, k, 2}}\right)z_{i,k}&\nonumber \\
&-z_{i, k-1}+\sum_{l=1}^{k-1} \xi_{k-1, l}^{i} x_{i, l+1},\,k=2,\cdots,n_i& \label{eq:7}\\
u_{i}=\,&\alpha_{i, n_{i}}-\varphi_{i}^{T} \left(x_{i}\right) \hat{\theta}_{i}-\psi_i\left(x_{i}\right)
&\label{eq:8}
\end{flalign}
where $c_{i,k}$, $\varpi_{ij,k,1}$ and $\varpi_{ij,k,2}$, $k=1,\cdots,n_i$ are positive design parameters, $\xi_{k-1, l}^{i}(k=2,\cdots,n_i,l=1,\cdots,k)$ is the partial derivative of $\alpha_{i,k-1}$ to $x_{i,l}$, which is a constant that relies on $c_{i,k}$, $\varpi_{ij,l,1}$ and $\varpi_{ij, l, 2}$. The updating law of ${{\hat \theta}_{i}}$ is designed as:
\begin{flalign}
&\dot{\hat{\theta}}_{i}=\Gamma_{i}
\left[-\sigma_{i}\hat{\theta}_{i}+\varphi_{i}\left({x}_{i}\right) {z}_{i, n_{i}}\right]&\label{eq:9}
\end{flalign}
where $\sigma_{i}$ is some positive design parameter, $\hat{\theta}_{i}$ is the estimate of ${\theta}_{i}$, with ${\tilde \theta}_{i}={\theta}_{i}-
\hat{\theta}_{i}$, and $\Gamma _{i}$ is a positive definite design matrix.

At this stage, the following lemma is introduced.

${\textbf{Lemma 1}}$. \cite{cai2022decentralized}
The state vector $x_i$ and its transformation vector $z_i$ obey the following relationship, with $z_i=[z_{i,1},\cdots,z_{i,n_i}]^{T}$:
\begin{flalign}
&\left\|x_{i}\right\| \leq\left\|A_{i}^{-1} B_{i}\right\|_{F} \left\|z_{i}\right\|&\label{eq:07289}
\end{flalign}
where $A_{i}$ and $B_{i}$ are constant matrices defined in (\ref{eq:66}) and (\ref{eq:67}).

{\textbf{Proof}}. See \emph{Appendix A}.

Now we are ready to state the following theorem.

${\textbf{Theorem 1}}$.
Consider the interconnected nonlinear non-triangular system (\ref{eq:1}) under \emph{Assumptions} 1-3, if using the decentralized adaptive controller (\ref{eq:8}), with the adaptive law (\ref{eq:9}), then it holds that: i) the global uniform boundedness of the closed-loop signals is ensured; and ii) all the subsystem outputs are steered into a residual set around zero, and the stabilization performance can be improved with some proper choices of the design parameters.

${\textbf{Proof}}$. The proof is composed of the following $n_i$ steps.

$\textbf{Step}$ 1:
Consider the Lyapunov function ${V_{i,1}} = \frac{1}{2}z_{i,1}^2$. From (\ref{eq:1}), (\ref{eq:4}) and (\ref{eq:5}), the derivative of ${V_{i,1}}$ is computed as
\begin{flalign}
&{\dot V_{i,1}} =z_{i,1}\left({z_{i,2}} + {\alpha _{i,1}}\right) +z_{i,1}f_{ii,1}+z_{i,1}\sum_{j\neq i} f_{ij,1}.& \label{eq:24}
\end{flalign}
According to \emph{Assumption} 1, it is derived that
\begin{flalign}
\left| z_{i,1} f_{ii,1}\right| \leq\, &\frac{1}{4 \varpi_{i i, 1,1}}z_{i, 1}^{2}+\varpi_{ii, 1,1}\hbar_{ii,1}^2\left\|x_{i}\right\|^{2}& \nonumber\\
&+\frac{1}{4 \varpi_{ii,1,2}}z_{i, 1}^{2}+\varpi_{i i, 1,2} \epsilon_{i i, 1}^{2} & \label{eq:25}\\
\left|z_{i,1} \sum_{j \neq i}f_{ij,1}\right| \leq \, &\sum_{j \neq i}\left(\frac{1}{4\varpi_{i j, 1,1}}z_{i, 1}^{2}+\,\varpi_{i j, 1,1}\hbar_{i j, 1}^2\left\|x_{j}\right\|^{2}\right.& \nonumber\\
&\left.+\frac{1}{4 \varpi_{i j, 1,2}}z_{i, 1}^{2}+\varpi_{i j, 1,2} \epsilon_{i j, 1}^{2}\right).& \label{eq:l25}
\end{flalign}
By utilizing (\ref{eq:6}), (\ref{eq:25}) and (\ref{eq:l25}), $\dot{V}_{i,1}$ is expressed as
\begin{flalign}
\dot{V}_{i, 1} \leq &-c_{i, 1} z_{i, 1}^{2}+z_{i, 1} z_{i, 2}+\varpi_{i i, 1,1}\hbar_{i i,1}^2\left\|x_{i}\right\|^{2}+\varpi_{ii,1,2} \epsilon_{i i, 1}^{2}& \nonumber\\
&+\sum_{j \neq i}\varpi_{ij,1,1} \hbar_{i j, 1}^2\left\|x_{j}\right\|^{2}+\sum_{j \neq i}\varpi_{i j, 1,2} \epsilon_{i j, 1}^{2}.& \label{eq:26}
\end{flalign}
%where the terms $\varpi_{i i, 1,1}\hbar_{i i,1}^2\left\|x_{i}\right\|^{2}$ and $\sum_{j \neq i}\varpi_{ij,1,1} \hbar_{i j, 1}^2\left\|x_{j}\right\|^{2}$ will be handled in the final step.

$\textbf{Step \emph{$k$}}$ $(k=2,\cdots,n_i-1)$:
Consider the Lyapunov function ${V_{i,k}} = {V_{i,k - 1}} + \frac{1}{2}z_{i,k}^2$. Using (\ref{eq:1}) and (\ref{eq:5}), $\dot{V}_{i,k}$ is derived as
\begin{flalign}
\dot{V}_{i, k}=\,&\dot{V}_{i, {k-1}}+z_{i,k}(z_{i, k+1}+\alpha_{i, k})+z_{i,k}f_{ii, k}+z_{i,k}\sum_{j \neq i} f_{ij,k}&\nonumber\\
&-z_{i,k}\sum_{l=1}^{k-1} \xi_{k-1, l}^{i}\left(x_{i, l+1}+f_{i i, l}+\sum_{j \neq i} f_{i j, l}\right)& \label{eq:28}
\end{flalign}
Based on \emph{Assumption} 1 and using Young's  inequality, it is seen that
\begin{flalign}
\left|z_{i, k} f_{i i, k}\right| \leq \, & \frac{1}{4 \varpi_{i i, k, 1}}z_{i, k}^{2}+\varpi_{ii,k, 1}\hbar_{i i, k}^{2}\left\|x_{i}\right\|^{2}&\nonumber\\
&+\frac{1}{4\varpi_{ii,k,2}}z_{i,k}^{2}+\varpi_{ii,k,2} \epsilon_{ii,k}^{2}& \label{eq:29}\\
\left|z_{i, k} \sum_{j \neq i} f_{i j, k}\right| \leq \,& \sum_{j \neq i}\left(\frac{1}{4 \varpi_{i j, k, 1}}z_{i, k}^{2}+\varpi_{i j, k, 1}\hbar_{i j, k}^2\left\|x_{j}\right\|^{2}\right.&\nonumber\\
&\left.+\,\frac{1}{4 \varpi_{ij,k,2}}z_{i, k}^{2}+\varpi_{i j, k, 2} \epsilon_{i j, k}^{2}\right)& \label{eq:30}
\end{flalign}
\begin{flalign}
&\left|z_{i, k} \sum_{l=1}^{k-1} \xi_{k-1,1}^{i} f_{i i, l}\right|
\leq \, \sum_{l=1}^{k-1}\left( \frac{\left(\xi_{k-1, l}^{i}\right)^{2}}{4 \varpi_{i i, k, 1}}z_{i, k}^{2}\right.&\nonumber\\
&\left.+\,\varpi_{ii,k,1}\hbar_{ii, l}^2
\left\|x_{i}\right\|^{2}+ \frac{\left(\xi_{k-1, l}^{i}\right)^{2}}{4 \varpi_{i i, k, 2}}z_{i, k}^{2}+\varpi_{i i, k, 2} \epsilon_{i i, l}^{2}\right)& \label{eq:31}\\
&\left|z_{i, k} \sum_{l=1}^{k-1} \xi_{k-1, l}^{i} \sum_{j \neq i} f_{i j, l}\right|
\leq \, \sum_{j \neq i} \sum_{l=1}^{k-1}\left(\frac{\left(\xi_{k-1, l}^{i}\right)^{2}}{4 \varpi_{i j, k, 1}}z_{i, k}^{2} \right. &\nonumber\\
&\left.+\,\varpi_{i j, k, 1}\hbar_{i j, l}^2\left\|x_{j}\right\|^{2}{\rm{+}} \frac{\left(\xi_{k-1, l}^{i}\right)^{2}}{4 \varpi_{i j, k, 2}}z_{i, k}^{2} +\varpi_{i j, k, 2} \epsilon_{i j, l}^{2}\right).& \label{eq:32}
\end{flalign}
By using (\ref{eq:7}), (\ref{eq:26}), (\ref{eq:29})-(\ref{eq:32}), it can be derived from (\ref{eq:28}) that
\begin{flalign}
\dot{V}_{i, k} \leq & -\sum_{\tau=1}^{k} c_{i, \tau}z_{i, \tau}^{2}+z_{i, k} z_{i, k+1}+\sum_{\tau=1}^{k} \sum_{l=1}^{\tau} \sum_{j=1}^{N}\left(\varpi_{i j, \tau, 1}\right. &\nonumber\\
&\left.\hbar_{i j, l}^2\left\|x_{j}\right\|^{2}
+\varpi_{i j, \tau, 2} \epsilon_{i j, l}^{2}\right).& \label{eq:33}
\end{flalign}
%where the term $\sum_{\tau=1}^{k} \sum_{l=1}^{\tau} \sum_{j=1}^{N}\left(\varpi_{i j, \tau, 1}\hbar_{i j, l}^2\left\|x_{j}\right\|^{2}+\varpi_{i j, \tau, 2} \epsilon_{i j, l}^{2}\right)$ will be coped with in the final step.

$\textbf{Step \emph{$n_i$}}$:
Consider the following Lyapunov function:
\begin{flalign}
&V_{i,n_i} = V_{i,n_i - 1} +\frac{1}{2}z_{i,n_i}^2 +\frac{1}{2}({k_{\theta,i}-\hat\theta_i})^T
\Gamma^{-1}_{i}({k_{\theta,i}-\hat\theta_i}) & \label{eq:33}
\end{flalign}
where $k_{\theta,i}$ is an unknown bounded constant vector. Different from the related control designs \cite{xing2016event,wang2020adaptive,wang2021adaptive},  here we construct the adaptive parameter term $\frac{1}{2}({k_{\theta,i}-\hat\theta_i})^T\Gamma^{-1}_{i}({k_{\theta,i}-\hat\theta_i})$, instead of $\frac{1}{2}({{\theta_i}-\hat\theta_i})^T
\Gamma^{-1}_{i}({{\theta_i}-\hat\theta_i})$, which is one of the key steps to avoid the appearance of $\dot{\theta}_i$, while ensuring system stability simultaneously, as detailed in the sequel.
From (\ref{eq:1}), (\ref{eq:5}), (\ref{eq:8}),  (\ref{eq:33}) and using Young's inequality, $\dot{V}_{i,n_i}$ is evaluated as
\begin{flalign}
\dot{V}_{i,n_i} {\rm{\leq}} & -\sum_{\tau=1}^{n_i} c_{i, \tau}z_{i, \tau}^{2}+\sum_{\tau=1}^{n_i-1} \sum_{l=1}^{\tau} \sum_{j=1}^{N}\left(\varpi_{i j, \tau, 1}\hbar_{i j, l}^2
\left\|x_{j}\right\|^{2}\right. &\nonumber\\
&\left.+\,\varpi_{i j, \tau, 2} \epsilon_{i j, l}^{2}\right){\rm{-}}({k_{\theta,i}{\rm{-}}\hat\theta_i})^T
\Gamma^{-1}_{i}\dot{\hat\theta}_i+\varphi_{i}^{T}
(x_{i})\tilde{\theta}_i z_{i,n_i} &\nonumber\\
&+\sum_{j=1}^{N}\left(\varpi_{i j, n_i, 1}\hbar_{i j,n_i}^2\left\|x_{j}\right\|^{2}
+\varpi_{i j, n_i, 2} \epsilon_{i j, n_i}^{2}\right)&\nonumber\\
&+\sum_{j=1}^{N}\sum_{l=1}^{n_i-1}\left(\varpi_{i j, n_i, 1}\hbar_{i j, l}^2\left\|x_{j}\right\|^{2}
+\varpi_{i j, n_i, 2}\epsilon_{i j, l}^{2}\right)& \label{eq:53}
\end{flalign}
Substituting (\ref{eq:9}) into (\ref{eq:53}) yields
\begin{flalign}
\dot{V}_{i,n_i}\leq& -\sum_{\tau=1}^{n_i} c_{i, \tau}z_{i, \tau}^{2}+\sum_{\tau=1}^{n_i} \sum_{l=1}^{\tau} \sum_{j=1}^{N}\left(\varpi_{i j, \tau, 1}\hbar_{i j,l}^2\left\|x_{j}\right\|^{2}\right. &\nonumber\\
&\left.+\,\varpi_{i j, \tau, 2} \epsilon_{i j, l}^{2}\right)-\frac{\sigma_{i}}{2}\left({k_{\theta,i}-\hat\theta_i}\right)^T
\left({k_{\theta,i}-\hat\theta_i}\right)&\nonumber\\
&+\varphi_{i}^T\left({x}_{i}\right)\Delta_{\theta_i}{z}_{i,n_{i}} +\frac{\sigma_{i}}{2}\left\|k_{\theta,i}\right\|^{2}\label{eq:f54}
\end{flalign}
where $\Delta_{\theta_i}={\theta_{i}}-{k_{\theta,i}}$.
In accordance with \emph{Lemma} 1 and \emph{Assumptions} 1-2, it can be obtained that
\begin{flalign}
\left|\varphi_{i}^T({x}_{i})
\Delta_{\theta_i}{z}_{i,n_{i}}\right|\le\, & \frac{1}{2}{\beta_{{\theta}_i}}L_{\varphi_{i}}^2 \left\|{x}_{i}\right\|^2+\frac{1}{2}{\beta_{{\theta}_i}}
\left\|{z}_{i}\right\|^2&\nonumber\\
\le\, &\frac{1}{2}{\beta_{{\theta}_i}}
\left(L_{\varphi_{i}}^2\left\|A_{i}^{-1}B_{i}\right\|_{F}^2
+1\right)\left\|z_{i}\right\|^2
&\label{eq:44}
\end{flalign}
Here, we pause to stress that, the effect of the parameter-induced perturbation term $\left|\varphi_{i}^T({x}_{i})
\Delta_{\theta_i}{z}_{i,n_{i}}\right|$ is handled in a non-compensatory manner due to the involvement of ETM, rather than making compensation for it as proposed in \cite{chen2020adaptive}, see \emph{Remark} 7 for more details. By using (\ref{eq:44}), then $\dot{V}_{i,n_i}$ can be further bounded as
\begin{flalign}
\dot{V}_{i,n_i} \leq & -\sum_{\tau=1}^{n_i} c_{i, \tau}z_{i, \tau}^{2}+\frac{1}{2}{\beta_{{\theta}_i}}
\left(L_{\varphi_{i}}^2\left\|A_{i}^{-1}B_{i}\right\|_{F}^2
+1\right)\left\|z_{i}\right\|^2 &\nonumber\\
&+\sum_{\tau=1}^{n_i} \sum_{l=1}^{\tau} \sum_{j=1}^{N}\left(\varpi_{i j, \tau, 1}\hbar_{i j,l}^2\left\|x_{j}\right\|^{2}+\,\varpi_{i j, \tau, 2} \epsilon_{i j, l}^{2}\right) &\nonumber\\
&-\frac{\sigma_{i}}{2}\left({k_{\theta,i}-\hat\theta_i}\right)^T
\left({k_{\theta,i}-\hat\theta_i}\right) +\frac{\sigma_{i}}{2}
\left\|k_{\theta,i}\right\|^{2}.
& \label{eq:57}
\end{flalign}
Consider the Lyapunov function $V=\sum_{i=1}^{N} V_{i, n_{i}}$, we can obtain
\begin{flalign}
\dot{V} \leq & -\sum_{i=1}^{N} \underline{c}_{i} \left\|z_{i}\right\|^2+\sum_{i=1}^{N}\frac{1}{2}{\beta_{{\theta}_i}}\left(L_{\varphi_{i}}^2\left\|A_{i}^{-1}B_{i}\right\|_{F}^2
+1\right)\left\|z_{i}\right\|^2 &\nonumber\\
&+\sum_{i=1}^{N}\sum_{\tau=1}^{n_i} \sum_{l=1}^{\tau} \sum_{j=1}^{N}\left(\varpi_{i j, \tau, 1}\hbar_{i j,l}^2\left\|x_{j}\right\|^{2}+\, \varpi_{i j, \tau, 2} \epsilon_{i j, l}^{2}\right)&\nonumber\\
&-\sum_{i=1}^{N}\frac{\sigma_{i}}{2}\left({k_{\theta,i}{\rm{-}}\hat\theta_i}\right)^T
\left({k_{\theta,i}{\rm{-}}\hat\theta_i}\right)+\sum_{i=1}^{N}\frac{\sigma_{i}}{2} \left\|k_{\theta,i}\right\|^{2}.
& \label{eq:58}
\end{flalign}
where $\underline{c}_i=\min\{c_{i,1},\cdots,c_{i,n_i}\}$.
By applying \emph{Lemma} 1, it can be derived from (\ref{eq:58}) that
\begin{flalign}
\dot{V} \leq&-\sum_{j=1}^{N} \underline{c}_{j} \left\|z_{j}\right\|^2+\sum_{j=1}^{N}\frac{1}{2}{\beta_{{\theta}_j}}\left(L_{\varphi_{j}}^2\left\|A_{j}^{-1}B_{j}\right\|_{F}^2+1\right)\left\|z_{j}\right\|^2&\nonumber\\
&+\sum_{j=1}^{N}\left(\sum_{\tau=1}^{n_i} \sum_{l=1}^{\tau} \sum_{i=1}^{N}\varpi_{i j, \tau, 1}\hbar_{ij,l}^2\left\|A_{j}^{-1}B_{j}\right\|_{F}^2\right)
\left\|z_{j}\right\|^{2} &\nonumber\\
&-\sum_{i=1}^{N}\frac{\sigma_{i}}{2}\left({k_{\theta,i}-\hat\theta_i}\right)^T
\left({k_{\theta,i}-\hat\theta_i}\right)+\Delta
& \nonumber\\
\le & -\sum_{j=1}^{N}{c}_{j}^*
\left\|z_{j}\right\|^{2}{\rm{-}}\sum_{i=1}^{N}\sigma_{i}^*
\left({k_{\theta,i}{\rm{-}}\hat\theta_i}\right)^T\left({k_{\theta,i}{\rm{-}}\hat\theta_i}\right)
+\Delta& \label{eq:61}
\end{flalign}
where ${c}_{j}^*=\underline{c}_{j}
-\sum_{\tau=1}^{n_i}\sum_{l=1}^{\tau} \sum_{i=1}^{N}\varpi_{i j,\tau, 1}\hbar_{ij,l}^2
\left\|A_{j}^{-1}B_{j}\right\|_{F}^2-\frac{1}{2}
{\beta_{{\theta}_j}}L_{\varphi_{j}}^2\left\|A_{j}^{-1}B_{j}
\right\|_{F}^2-\frac{1}{2}{\beta_{{\theta}_j}}>0$ by choosing $\underline{c}_{j}$, i.e., $\min\{c_{j,1},\cdots,c_{j,n_i}\}$ larger enough, $\sigma_{i}^*=\frac{\sigma_{i}}{2}>0$, and $\Delta=\sum_{i=1}^{N}\sum_{\tau=1}^{n_i} \sum_{l=1}^{\tau} \sum_{j=1}^{N}\varpi_{i j, \tau, 2} \epsilon_{i,j,l}^{2}+\sum_{i=1}^{N}\frac{\sigma_{i}}{2}
\left\|k_{\theta,i}\right\|^{2}$.
%Similar to \cite{cai2022decentralized}, , it can be ensured that ${c}_{j}^*>0$ and $\sigma_{i}^*>0$.
Then it holds that $\dot{V} \le -lV+\Delta$, with  $l=\min\left \{2{c}_{1}^*,\cdots,2{c}_{N}^*,
\frac{2{\sigma}_{1}^*}{\lambda_{\max}\{\Gamma_1^{-1}\}},\cdots,
\frac{2{\sigma}_{N}^*}{\lambda_{\max}\{\Gamma_N^{-1}\}}\right\}$.

Now we are ready to prove in detail that the results i) and ii) in \emph{Theorem} 1 are ensured.

$\bullet$ \emph{Stability Analysis.}
From the above analysis, we have ${V}\left( t \right) \le {e^{-l t}}{V}\left( 0 \right) + \frac{{\Delta}}{l}\left( {1 - {e^{-l t}}} \right)\in {\mathcal{L}_\infty}$, it follows that ${z_{i,k}}$ and ${\tilde{\theta}_{i}}$ are bounded, $k = 1, \cdots ,n_i$. From (\ref{eq:4}), (\ref{eq:5}), (\ref{eq:6}) and (\ref{eq:7}), it is established that ${x_{i,k}}$ is bounded, $k=1,\cdots,n_i$. Then it can be derived from (\ref{eq:8}) that $u_i$ is bounded. Therefore, all signals in the closed-loop system are globally uniformly  bounded.

$\bullet$ \emph{Performance Analysis.}
As $\dot{V} \le -lV+\Delta$, we have $\left| {z_{i,1}} \right|\le \sqrt {2V}  = \sqrt {2\left({V\left( 0 \right) - \frac{\Delta}{l}} \right){e^{-l t}} + 2\frac{\Delta}{l}}$, which implies that $z_{i,1}$ is ensured to attenuate to a residual set around zero. In addition, from the definition of $V$, $l$ and $\Delta$, it holds that the upper bound of $\left| {z_{i,1}} \right| $ can be decreased by increasing design parameters $c_{i,k}$ and $\Gamma_i$, or decreasing design parameter $\varpi_{i j, k,1}$ and $\varpi_{i j, k,2}$, $k=1,\cdots,n_i$.   $\hfill{\blacksquare}$

${\textbf{Remark 3}}$.
The non-triangular structural uncertainties involve both modeling errors and nonlinear coupling interactions, thus are difficult to tackle. For the first part, by constructing a nonlinear compensation term $-\sum_{j\neq i}^{N}\left(\frac{1}{4 \varpi_{i j, k, 1}}+\frac{1}{4 \varpi_{i j, k,2}}\right)\bar{z}_{i, k}-\sum_{j\neq i}^{N}\sum_{l=1}^{k-1}\left
(\frac{\left(\xi_{k-1, l}^{i}\right)^{2}}{4 \varpi_{i j, k,1}}+\frac{\left(\xi_{k-1, l}^{i}\right)^{2}}{4 \varpi_{i j, k, 2}}\right)\bar{z}_{i,k}$ in the virtual controllers of each subsystem $\alpha_{i,k-1} (k=2,\cdots,n_i)$, as seen in (\ref{eq:7}), we naturally offset the terms related to $z_{i,k}$ in (\ref{eq:30}) and (\ref{eq:32}). Whereas the uncertainties in the second part contain nonlinear coupling interactions among subsystems, it is even more challenging to handle. Here, inspired by the ideas in \cite{cai2022decentralized}, we first keep all terms associated with $\left\|x_j\right\|^2(j=1,\cdots,N)$ in each recursive step, and then cope with them in the final step with the aid of \emph{Lemma} 1. %via designing appropriate control laws.
It is worth noting that what is considered here is an entirely different and more difficult implementation scenario than the one in \cite{cai2022decentralized}, where the traditional backstepping method can be directly used in \cite{cai2022decentralized} as state/input-triggering is not considered. Meanwhile the work in \cite{cai2022decentralized} involves only constant parametric uncertainties  which therefore can be easily handled by using adaptive parameter estimates methods.

\section{Decentralized Event-triggered Adaptive Backstepping Control}
In this section, a decentralized adaptive backstepping control scheme under event-triggering setting is constructed upon the previous scheme, with feasibility and stability analysis provided. Such strategy not only inherits the ability of handling non-triangular structural uncertainties and time-varying parameters in the continuous scheme, but also evades the non-differentiability issue.

\subsection{Event Triggering Mechanism}
Inspired by the ETM presented in \cite{xing2016event,wang2020adaptive},
we denote ${{\bar x}_{i,k}}$, ${\bar x}_{j,k}$ and $u_i$, $i,j=1,\cdots,N\,(j\neq i)$, $k=1,\cdots,n_i$ as the local states information, other subsystem states information and the actuation signal information, respectively,  which broadcast their information according to the devised ETM.
Since $t_{k,l}^{i}$, $t_{k,l}^{j}$ and $t_{u,l}^{i}$ denote the $l$th event time for system $i$, other subsystem $j$ and actuation signal broadcasting theirs information, respectively, which indicates that ${{\bar x}_{i,k}}$, ${{\bar x}_{j,k}}$ and $u_i$ remain  unchanged as
\begin{flalign}
{{\bar x}_{i,k}}\left( t \right) =\,&{x_{i,k}}\left( {t_{k,l}^{i}} \right),\, \forall t \in [t_{k,l}^{i},t_{k,l + 1}^{i}) &\nonumber \\
{{\bar x}_{j,k}}\left( t \right) =\,&{x_{j,k}}\left( {t_{k,l}^{j}} \right),\, \forall t \in [t_{k,l}^{j},t_{k,l + 1}^{j})&\nonumber \\
{u_i}\left( t \right) =\,&{v_i}\left( {t_{u,l}^{i}} \right),\,\forall t \in [t_{u,l}^{i},t_{u,l + 1}^{i})\label{eq:aa10}
\end{flalign}
for $l = 0,1,2, \cdots$.
Now we propose the following triggering conditions that only depends on locally available information:
\begin{flalign}
t_{k,l + 1}^i=& \inf \left\{ {t > t_{k,l}^i,\left| {{x_{i,k}}\left( t \right) - {{\bar x}_{i,k}}\left( t \right)} \right| > \Delta x_{i,k}} \right\} & \label{eq:10}\\
t_{k,l + 1}^j=& \inf \left\{ {t > t_{k,l}^j,\left| {{x_{j,k}}\left( t \right) - {{\bar x}_{j,k}}\left( t \right)} \right| > \Delta x_{j,k}} \right\} & \label{eq:f10}\\
t_{{u},l + 1}^i=& \inf \left\{ {t > {t_{{u},l}^i},\left| {v_i\left( t \right) - u_i\left( t \right)} \right| > \Delta u_i} \right\} & \label{eq:11}
\end{flalign}
where $\Delta x_{i,k}$, $\Delta x_{j,k}$ and $\Delta{u_i}$ are positive triggering thresholds, $t_{k,0}^i$, $t_{k,0}^j$ and $t_{{u},0}^i$ denote the first instants when (\ref{eq:10})-(\ref{eq:11}) are fulfilled, respectively.

${\textbf{Remark 4}}$.
%The result here deals with the scenario where both the states sensoring and actuation signal transmission in each subsystem are executed intermittently on the event-driven basis.
The designed ETM allows all the states sensoring and data transmission to be executed intermittently on the event-driven basis, and the states include those from the subsystem itself and its neighbors (not just the states between subsystems), thus different from that in \cite{wang2020adaptive,wang2021adaptive}, which implies that the sensors do not need to be powered all the time and the data from the sensors to the controllers does not have to be transmitted ceaselessly. Besides, the communication between the control unit and the system can be made less frequently. In such a way, the proposed approach is more efficient in terms of saving communication and energy resources (despite the systems under consideration are more general interconnected nonlinear non-triangular forms) in comparison to the existing ones \cite{xing2016event,ghodrat2018local,wang2020adaptive,wang2021adaptive}, therein either states or control input are transmitted intermittently over the network.

\begin{figure}
\begin{center}
\includegraphics[width=0.48\textwidth,height=40mm]{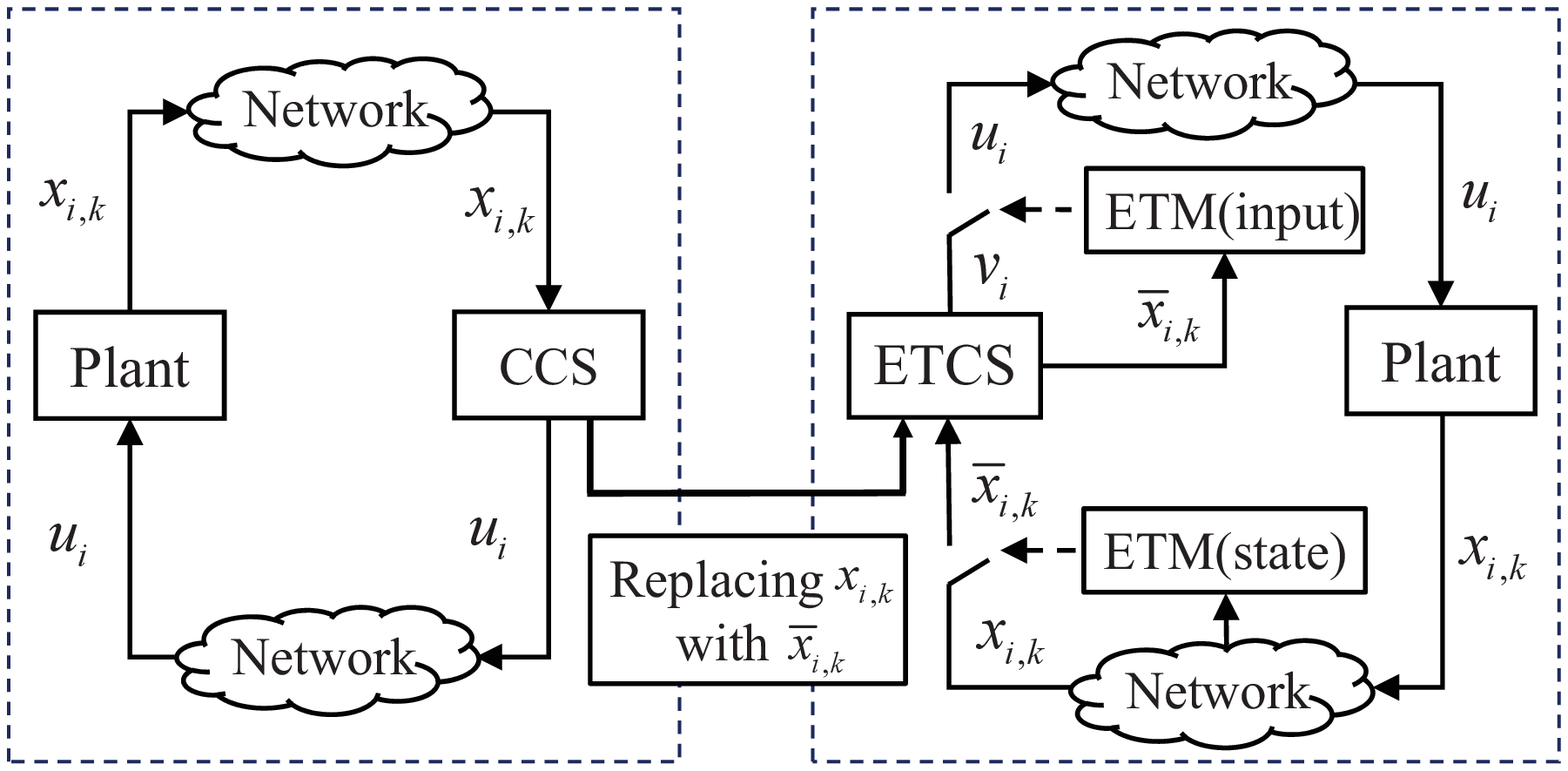}
\caption{The block diagram of closed-loop systems with continuous control scheme (CCS) and the corresponding event-triggered control scheme (ETCS).}
\end{center}
\end{figure}

\subsection{Controller Design}
Since only locally intermittent state signals $\bar{x}_{i,k}$ (rather than its continuous value) are available in controlling the system in case of  event-triggering, we modify the coordinate transformation defined in (\ref{eq:4})-(\ref{eq:5}) into the following form by replacing $x_{i,k}$ with $\bar{x}_{i,k}$:
\begin{flalign}
{{\bar z}_{i,1}} =\,& {{\bar x}_{i,1}}& \label{eq:12}\\
{{\bar z}_{i,k}} =\,& {{\bar x}_{i,k}} - {{\bar \alpha} _{i,k-1}},k = 2, \cdots ,n_i & \label{eq:13}
\end{flalign}

Based upon intermittent state feedback, the decentralized event-triggered adaptive backstepping control scheme is constructed as:
\begin{flalign}
{\bar \alpha}_{i, 1}=&-c_{i, 1} {\bar{z}}_{i, 1}-\frac{1}{4 \varpi_{i i, 1,1}} {\bar{z}}_{i, 1}-\frac{1}{4 \varpi_{i i, 1,2}} {\bar{z}}_{i, 1}&\nonumber \\
&-\sum_{j \neq i} \frac{1}{4 \varpi_{i j, 1,1}} {\bar{z}}_{i, 1}-\sum_{j \neq i}\frac{1}{4 \varpi_{i j, 1,2}} {\bar{z}}_{i, 1}& \label{eq:14}\\
\bar{\alpha}_{i, k}=&-c_{i, k} \bar{z}_{i, k}-\sum_{j=1}^{N}
\left(\frac{1}{4 \varpi_{i j, k, 1}}+\frac{1}{4 \varpi_{i j, k,2}}\right)\bar{z}_{i, k}&\nonumber \\
&-\sum_{j=1}^{N}\sum_{l=1}^{k-1}
\left(\frac{\left(\xi_{k-1, l}^{i}\right)^{2}}{4 \varpi_{i j, k, 1}}+\frac{\left(\xi_{k-1, l}^{i}\right)^{2}}{4 \varpi_{i j, k, 2}}\right)\bar{z}_{i,k}&\nonumber \\
&-\bar{z}_{i, k-1}+\sum_{l=1}^{k-1} \xi_{k-1, l}^{i} \bar{x}_{i, l+1},\,k=2,\cdots,n_i& \label{eq:15}\\
v_{i}=\,&{\bar \alpha}_{i, n_{i}}-\varphi_{i}^{T}\left({\bar x}_{i}\right) \hat{\theta}_{i}-\psi_i\left({\bar x}_{i}\right)
&\label{eq:16}
\end{flalign}
where $c_{i,k}$, $\varpi_{ij,k,1}$ and $\varpi_{ij,k,2}$, $k=1,\cdots,n_i$ are positive design parameters.
The updating laws of ${\hat{\theta}}_{i}$ is designed as:
\begin{flalign}
&\dot{\hat{\theta}}_{i}=\Gamma_{i}
\left[-\sigma_{i}\hat{\theta}_{i}+\varphi_{i}\left({\bar x}_{i}\right) {\bar z}_{i, n_{i}}\right]&\label{eq:17}
\end{flalign}
where ${\sigma_{i}}$ is a positive design parameter and $\Gamma_{i}$ is a positive definite design matrix.
The proposed two globally decentralized adaptive backstepping control strategies and their relationship are conceptually shown in Fig. 1.

To ensure the global uniform boundedness of all the closed-loop signals, we establish the following lemma.

${\textbf{Lemma 2}}$.
The effects of event-triggering are bounded as follows:
\begin{flalign}
\left| {{z_{i,k}}-{{\bar z}_{i,k}}}\right|\le\,&\Delta {z_{i,k}} &\label{eq:22}\\
\left|{\alpha_{i,k}{\rm{-}}{\bar \alpha}_{i,k}}\right|\le\, &\Delta {\alpha_{i,k}}&\label{eq:23}
\end{flalign}
for $i=1,\cdots,N$, $k=1,\cdots,n_i$, where $\Delta {z_{i,k}}$ and $\Delta {\alpha_{i,k}}$ are positive constants that depend on the triggering thresholds $\Delta x_{i,k}$, $\Delta x_{j,k}$ and $\Delta u_i$, and the design parameters $c_{i,k}$, $\varpi_{ij,k,1}$ and $\varpi_{ij,k,2}$.

{\textbf{Proof}}. See \emph{Appendix B}.

${\textbf{Remark 5}}$.
Thanks to the proposed modified congelation of variables based approach and a special treatment on non-triangular uncertainties, the partial derivatives $\xi_{k-1,l}^{i}\,(k=2,\cdots,n_i,l=1,\cdots,k)$ in each subsystem are all ensured to be constant. Such property ensures that the impacts of event-triggering are bounded by constants, as detailed in \emph{Lemma} 2. It is not trivial to derive such property, especially in the presence of serious uncertainties and time-varying parameters. Specifically, in the available adaptive state-triggered results such as \cite{wang2021adaptive,zhang2021adaptive,sun2022Distributed}, only systems in norm form exhibit this property \cite{wang2021adaptive};
in the nonlinear strict-feedback systems with parametric/non-parametric uncertainties  \cite{zhang2021adaptive,sun2022Distributed}, one can only prove that the triggering errors are bounded by some time-varying functions, while requiring the exploitation of DSC techniques or neural networks (NN) based approximators (which can only obtain a semi-global result).
Therefore it is even more challenging to retain such property for the non-triangular nonlinear time-varying interconnected systems actually considered here.

We are in the position to state the following theorem.

${\textbf{Theorem 2}}$.
Consider the interconnected nonlinear non-triangular system (\ref{eq:1}) under \emph{Assumptions} 1-3, if using the decentralized adaptive controller (\ref{eq:16}), with adaptive law (\ref{eq:17}) and triggering conditions (\ref{eq:10})-(\ref{eq:11}), it then holds that: i) the global uniform boundedness of the closed-loop signals is ensured; ii) all the subsystem outputs are steered into a residual set around zero, yet the stabilization performance can be improved with some proper choices of the design parameters; and iii) the Zeno phenomenon is precluded.
\vspace{0.1cm}

${\textbf{Proof}}$.
The proof of the claims in the theorem consists of two parts: stability analysis and exclusion of Zeno behavior.

${\textbf{1) Stability analysis}}$. This part consists of the following $n_i$ steps.\\
\indent
$\textbf{Step}$ 1:
Consider a Lyapunov function ${V_{i,1}} = \frac{1}{2}z_{i,1}^2$. From (\ref{eq:1}), (\ref{eq:4}), (\ref{eq:5}) and (\ref{eq:6}), the derivative of ${V}_{i,1}$ is expressed as
\begin{flalign}
\dot{V}_{i, 1} \leq &-c_{i, 1} z_{i, 1}^{2}+z_{i, 1} z_{i, 2}+\varpi_{i i, 1,1}\hbar_{i i,1}^2\left\|x_{i}\right\|^{2}+\varpi_{ii,1,2} \epsilon_{i i, 1}^{2}& \nonumber\\
&+\sum_{j \neq i}\varpi_{ij,1,1} \hbar_{i j, 1}^2\left\|x_{j}\right\|^{2}+\sum_{j \neq i}\varpi_{i j, 1,2} \epsilon_{i j, 1}^{2}.& \label{eq:51}
\end{flalign}
\indent
$\textbf{Step \emph{$k$}}$ $(k=2,\cdots,n_i-1)$:
Consider ${V_{i,k}} = {V_{i,k - 1}} + \frac{1}{2}z_{i,k}^2$. By using (\ref{eq:1}), (\ref{eq:5}), (\ref{eq:7}) and (\ref{eq:51}), we can deduce that
\begin{flalign}
\dot{V}_{i, k} \leq & -\sum_{\tau=1}^{k} c_{i, \tau}z_{i, \tau}^{2}+z_{i, k} z_{i, k+1}+\sum_{\tau=1}^{k} \sum_{l=1}^{\tau} \sum_{j=1}^{N}\left(\varpi_{i j, \tau, 1}\right. &\nonumber\\
&\left.\hbar_{i j, l}^2\left\|x_{j}\right\|^{2}
+\varpi_{i j, \tau, 2} \epsilon_{i j, l}^{2}\right).& \label{eq:52}
\end{flalign}
\indent
$\textbf{Step \emph{$n_i$}}$:
Consider the Lyapunov function ${V_{i,n_i}} = {V_{i,n_i - 1}} + \frac{1}{2}z_{i,n_i}^2 + \frac{1}{2}({k_{\theta,i}-\hat\theta_i})^T\Gamma^{-1}_{i}({k_{\theta,i}-\hat\theta_i})$, where $k_{\theta,i}$ is defined as before.
Note that the control law $v_i$ in (\ref{eq:16}) can be rewritten as
\begin{flalign}
v_i =\, &{\alpha}_{i, n_{i}}
-\varphi_{i}\left({x}_{i}\right)^{T} \hat{\theta}_{i}-\psi_i\left({x}_{i}\right)+
\left({\bar \alpha}_{i, n_{i}}-{\alpha}_{i, n_{i}}\right)&\nonumber\\
&+\left(\varphi_{i}^{T}\left({x}_{i}\right)
-\varphi_{i}^{T}\left({\bar x}_{i}\right)
\right)\hat{\theta}_{i}+\left(\psi_{i}\left({x}_{i}\right)-\psi_i\left({\bar x}_{i}\right)\right).&\label{eq:35}
\end{flalign}
From (\ref{eq:1}), (\ref{eq:5}), (\ref{eq:52}) and (\ref{eq:35}), ${\dot{V}_{i,n_i}}$ is expressed as
\begin{flalign}
\dot{V}_{i,n_i} \leq & -\sum_{\tau=1}^{n_i} c_{i, \tau}z_{i, \tau}^{2}+\sum_{\tau=1}^{n_i-1} \sum_{l=1}^{\tau} \sum_{j=1}^{N}\left(\varpi_{i j, \tau,1}\hbar_{i,j,l}^2\left\|x_{j}\right\|_{2}^{2}\right. &\nonumber\\
&\left.+\,\varpi_{ij,\tau, 2} \epsilon_{i j, l}^{2}\right){\rm{-}}({k_{\theta,i}{\rm{-}}\hat\theta_i})^T
\Gamma^{-1}_{i}\dot{\hat\theta}_i+
\varphi_{i}^{T}(x_{i}){\tilde{\theta}_{i}}z_{i,n_i}&\nonumber\\
&+\sum_{j=1}^{N}\left(\varpi_{i j, n_i, 1}^2\hbar_{i j,n_i}^2\left\|x_{j}\right\|^{2}
+\varpi_{i j, n_i, 2} \epsilon_{i j, n_i}^{2}\right)&\nonumber\\
&+\sum_{j=1}^{N}\sum_{l=1}^{n_i-1}\left(\varpi_{i j, n_i, 1}\hbar_{i j,l}^2\left\|x_{j}\right\|^{2}
+\varpi_{i j, n_i, 2} \epsilon_{i j, l}^{2}\right)&\nonumber\\
&+(u_i-v_i)z_{i,n_i}+\left(\varphi_{i}^{T}
\left({x}_{i}\right){\rm{-}}\varphi_{i}^{T}\left({\bar x}_{i}\right)\right)\hat{\theta}_{i}z_{i,{n_i}}&\nonumber\\
&+\left({\bar \alpha}_{i, n_{i}}
{\rm{-}}{\alpha}_{i,n_{i}}\right)z_{i,{n_i}}{\rm{+}}
\left(\psi_{i}({x}_{i}){\rm{-}}\psi_i({\bar x}_{i})\right)z_{i,{n_i}}. & \label{eq:36}
\end{flalign}
Substituting (\ref{eq:17}) into (\ref{eq:36}), it follows that
\begin{flalign}
\dot{V}_{i,n_i} \leq & -\sum_{\tau=1}^{n_i} c_{i, \tau}z_{i, \tau}^{2}+\sum_{\tau=1}^{n_i} \sum_{l=1}^{\tau} \sum_{j=1}^{N}\left(\varpi_{i j, \tau, 1}\hbar_{i j,l}^2\left\|x_{j}\right\|^{2}\right. &\nonumber\\
&\left.+\,\varpi_{i j, \tau, 2} \epsilon_{i j, l}^{2}\right)+\sigma_{i}({k_{\theta,i}
-\hat\theta_i})^T\hat{\theta}_{i}+(u_i-v_i)z_{i,n_i}&\nonumber\\
&+\left({\bar \alpha}_{i, n_{i}}
-{\alpha}_{i,n_{i}}\right)z_{i,{n_i}}
+\left(\psi_{i}\left({x}_{i}\right)-\psi_i\left({\bar x}_{i}\right)\right)z_{i,{n_i}}&\nonumber\\
&+\left(\varphi_{i}^{T}
\left({x}_{i}\right){\rm{-}}\varphi_{i}^{T}\left({\bar x}_{i}\right)\right)\hat{\theta}_{i}z_{i,{n_i}}+\varphi_{i}^T
\left({x}_{i}\right)\Delta_{\theta_i}{z}_{i,n_{i}} &\nonumber\\
&+\left({k_{\theta,i}-\hat\theta_i}\right)^T\left(\varphi_{i}
\left({x}_{i}\right){z}_{i, n_{i}}-\varphi_{i}\left({\bar x}_{i}\right){\bar z}_{i,n_{i}}\right)& \label{eq:d37}
\end{flalign}
where $\Delta_{\theta_i}={\theta_{i}}-{k_{\theta,i}}$.  Furthermore, $\dot{V}_{i,n_i}$ becomes
\begin{flalign}
\dot{V}_{i,n_i} \le& -\sum_{\tau=1}^{n_i} c_{i, \tau}z_{i, \tau}^{2}+\sum_{\tau=1}^{n_i} \sum_{l=1}^{\tau} \sum_{j=1}^{N}\left(\varpi_{i j, \tau, 1}\hbar_{i j,l}^2\left\|x_{j}\right\|^{2}\right. &\nonumber\\
&\left.+\,\varpi_{i j, \tau, 2} \epsilon_{i j, l}^{2}\right)-\frac{\sigma_{i}}{2}\left({k_{\theta,i}-\hat\theta_i}\right)^T
\left({k_{\theta,i}-\hat\theta_i}\right)&\nonumber\\
&+\Delta\Xi_i+\varphi_{i}^T\left({x}_{i}\right)\Delta_{\theta_i}{z}_{i,n_{i}}
+\frac{\sigma_{i}}{2}\left\|k_{\theta,i}\right\|^{2}. & \label{eq:40}
\end{flalign}
%where $\Delta\Xi_i=\left|u_i-v_i\right|\left|{z}_{i,n_{i}}\right| +\left|\left(\varphi_{i}^{T}\left({x}_{i}\right)-\varphi_{i}^{T}\left({\bar x}_{i}\right)\right)\hat{\theta}_{i}\right|\left|{z}_{i,n_{i}}\right|
%+\left|\left({k_{\theta,i}{\rm{-}}\hat\theta_i}\right)^T\left(\varphi_{i}\left({x}_{i}\right){z}_{i,n_{i}}{\rm{-}}\varphi_{i}\left({\bar x}_{i}\right){\bar z}_{i,n_{i}}\right)\right|+\left|\bar{\alpha}_{i,n_i}{\rm{-}}\alpha_{i,n_i}\right| \left|{z}_{i,n_{i}}\right|+\left|\psi_{i}\left({x}_{i}\right)-\psi_i\left({\bar x}_{i}\right)\right|\left|{z}_{i,n_{i}}\right|$.
with
\begin{flalign}
\Delta\Xi_i=&\left|u_i-v_i\right|\left|{z}_{i,n_{i}}\right| +\left|\left(\varphi_{i}^{T}\left({x}_{i}\right)-\varphi_{i}^{T}\left({\bar x}_{i}\right)\right)\hat{\theta}_{i}\right|\left|{z}_{i,n_{i}}\right| &\nonumber\\
&+\left|\bar{\alpha}_{i,n_i}-\alpha_{i,n_i}\right| \left|{z}_{i,n_{i}}\right|+\left|\psi_{i}\left({x}_{i}\right)-\psi_i\left({\bar x}_{i}\right)\right|\left|{z}_{i,n_{i}}\right| &\nonumber\\
&+\left|\left({k_{\theta,i}-\hat\theta_i}\right)^T\left(\varphi_{i}\left({x}_{i}\right){z}_{i,n_{i}}-\varphi_{i}\left({\bar x}_{i}\right){\bar z}_{i,n_{i}}\right)\right|.
&\label{eq:a37}
\end{flalign}
According to \emph{Assumption} 3, we can obtain that
\begin{flalign}
&\left|\left(\varphi_{i}^{T}
\left({x}_{i}\right)-\varphi_{i}^{T}\left({\bar x}_{i}\right)\right)\hat{\theta}_{i}{z_{i,n_i}}\right|&\nonumber\\
&\le \left\|\varphi_{i}^{T}\left({x}_{i}\right)
-\varphi_{i}^{T}\left({\bar x}_{i}\right)\right\| \left\|\hat{\theta}_{i}-k_{\theta,i}+k_{\theta,i} \right\|\left|{z_{i,n_i}}\right|&\nonumber\\
&\le L_{\varphi_{i}}\Delta x_i \left\|k_{\theta,i}-\hat{\theta}_{i}
\right\|\left|{z_{i,n_i}}\right|+
L_{\varphi_{i}}\Delta x_i \left\|k_{\theta,i}\right\|\left|{z_{i,n_i}}\right|&\label{eq:41}\\
&\left\|\varphi_{i}\left({x}_{i}\right){z}_{i, n_{i}}-\varphi_{i}\left({\bar x}_{i}\right){\bar z}_{i,n_{i}}\right\| &\nonumber\\
&\le \left\|\varphi_{i}\left({x}_{i}\right)
\right\| \left|{z}_{i,n_{i}}-\bar{z}_{i,n_{i}}\right| +\left\|\varphi_{i}\left({x}_{i}\right)-\varphi_{i}\left({\bar{x}}_{i}
\right)\right\|\left|{\bar{z}}_{i,n_{i}}\right|&\nonumber\\
&\le L_{\varphi_{i}} \Delta{z}_{i, n_{i}}\left\|{x}_{i} \right\|+L_{\varphi_{i}} \Delta{x}_{i}\left|{{z}}_{i, n_{i}}\right|+L_{\varphi_{i}}\Delta{x}_{i}\Delta{z}_{i, n_{i}}.&\label{eq:l41}
\end{flalign}
Notice from (\ref{eq:a37}), (\ref{eq:41}), (\ref{eq:l41}) and invoking \emph{Lemma} 1, it holds that
\begin{flalign}
\Delta\Xi_i%\le \,& \left(\Delta u_i+\Delta \alpha_{i,n_i}+L_{\psi_{i}}\Delta{x}_{i}+L_{\varphi_{i}}\Delta x_i\left\|k_{\theta,i}\right\|\right)\left\|{z_{i}}\right\|&\nonumber\\
%&+\left\|{k_{\theta,i}-\hat\theta_i}\right\|\left(L_{\varphi_{i}}\left\|A_{i}^{-1} B_{i}\right\|_{F}\Delta{z}_{i,n_{i}}\left\|z_{i}\right\|\right. &\nonumber\\
%&\left.+\,2L_{\varphi_{i}}\Delta{x}_{i}\left\|z_{i}\right\|
%+L_{\varphi_{i}} \Delta{x}_{i}\Delta{z}_{i, n_{i}}\right)&\nonumber\\
\le\,& \lambda_{i,1}\left\|z_{i}\right\|+\left\|{k_{\theta,i}
-\hat\theta_i}\right\|\left(\delta_{i,1} \left\|{z}_{i}\right\| +\delta_{i,2}\right)
&\nonumber\\
\le\,& \lambda_{i}\left\|z_{i}\right\|^2
+\delta_{i}\left({k_{\theta,i}-\hat\theta_i}\right)^T
\left({k_{\theta,i}-\hat\theta_i}\right)
+\Delta_{i,0}
&\label{eq:43}
\end{flalign}
where $\lambda_{i,1}=\Delta \alpha_{i,n_i}+ \Delta u_i+L_{\psi_{i}}\Delta{x}_{i}+L_{\varphi_{i}}\Delta x_i\left\|k_{\theta,i}\right\|$, $\delta_{i,1}=L_{\varphi_{i}}\left\|A_{i}^{-1} B_{i}\right\|_{F}\Delta{z}_{i,n_{i}}+2L_{\varphi_{i}} \Delta{x}_{i}$, $\delta_{i,2}=L_{\varphi_{i}} \Delta{x}_{i}\Delta{z}_{i, n_{i}}$,  $\lambda_{i}=\frac{1}{2}(\lambda_{i,1}+\delta_{i,1})$, $\delta_{i}=\frac{1}{2}(\delta_{i,1}
+\delta_{i,2})$ and $\Delta_{i,0}=\frac{1}{2}
(\lambda_{i,1}+\delta_{i,2})$.
In accordance with (\ref{eq:44}), (\ref{eq:40}) and (\ref{eq:43}), the following inequality holds
\begin{flalign}
\dot{V}_{i,n_i} \leq & -\sum_{\tau=1}^{n_i} c_{i, \tau}z_{i,\tau}^{2}+\frac{1}{2}{\beta_{{\theta}_i}}
\left(L_{\varphi_{i}}^2\left\|A_{i}^{-1}B_{i}\right\|_{F}^2
+1\right)\left\|z_{i}\right\|^2 &\nonumber\\
&+\sum_{\tau=1}^{n_i} \sum_{l=1}^{\tau} \sum_{j=1}^{N}\varpi_{i j, \tau, 1}^2\hbar_{i j,l}^2\left\|A_{j}^{-1} B_{j}\right\|_{F}^2\left\|z_{j}\right\|^2
{\rm{+}}\lambda_{i}\left\|z_{i}\right\|^2&\nonumber\\
&-\left(\frac{\sigma_{i}}{2}-\delta_{i}\right)\left({k_{\theta,i}-\hat\theta_i}\right)^T
\left({k_{\theta,i}-\hat\theta_i}\right)+\Delta_{i}& \label{eq:45}
\end{flalign}
where $\Delta_i=\sum_{\tau=1}^{n_i} \sum_{l=1}^{\tau} \sum_{j=1}^{N}\varpi_{i j, \tau, 2} \epsilon_{i j, l}^{2}+\frac{1}{2}\sigma_{i}\left\|k_{\theta,i}\right\|^{2}
+\Delta_{i,0}$.
Consider the Lyapunov function $V=\sum_{i=1}^{N} V_{i, n_{i}}$. With (\ref{eq:45}) in mind, we have
\begin{flalign}
\dot{V}\le&-\sum_{j=1}^{N} \underline{c}_j\left\|{z}_{j}\right\|^2
+\sum_{j=1}^{N}\frac{1}{2}{\beta_{{\theta}_j}}
\left(L_{\varphi_{j}}^2\left\|A_{j}^{-1}B_{j}\right\|_{F}^2
{\rm{+}}1\right)\left\|z_{j}\right\|^2&\nonumber\\
&{\rm{+}}\sum_{j=1}^{N}\left(\sum_{i=1}^{N}\sum_{\tau=1}^{n_i} \sum_{l=1}^{\tau}\varpi_{i j, \tau, 1}^2\hbar_{i j,l}^2\left\|A_{j}^{-1} B_{j}\right\|_{F}^2\right)\left\|z_{j}\right\|^2
&\nonumber\\
&+\sum_{j=1}^{N}\lambda_{j}\left\|z_{j}\right\|^2
{\rm{-}}\sum_{i=1}^{N}{\sigma}_{i}^*\left({k_{\theta,i}{\rm{-}}\hat\theta_i}\right)^T
\left({k_{\theta,i}{\rm{-}}\hat\theta_i}\right)+\Delta&\nonumber\\
\le & -\sum_{j=1}^{N}{c}_{j}^*
\left\|z_{j}\right\|^{2}{\rm{-}}\sum_{i=1}^{N}{\sigma}_{i}^*\left({k_{\theta,i}{\rm{-}}\hat\theta_i}\right)^T
\left({k_{\theta,i}{\rm{-}}\hat\theta_i}\right)+\Delta& \label{eq:47}
\end{flalign}
where $\underline{c}_j=\min\{c_{j,1},\cdots,c_{j,n_i}\}$, ${c}_{j}^*=\underline{c}_{j} -\sum_{i=1}^{N}\sum_{\tau=1}^{n_i} \sum_{l=1}^{\tau}\varpi_{i j, \tau, 1}\hbar_{ij,l}^2\left\|A_{j}^{-1}B_{j}\right\|_{F}^2
-\lambda_{j}-\frac{1}{2}{\beta_{{\theta}_j}}
\left(L_{\varphi_{j}}^2\left\|A_{j}^{-1}B_{j}\right\|_{F}^2
+1\right)>0$ and ${\sigma}_{i}^*=\frac{\sigma_{i}}{2}
-\delta_{i}>0$ by choosing $\underline{c}_{j}$, i.e., $\min\{c_{j,1},\cdots,c_{j,n_i}\}$ and $\sigma_{i}$ larger enough, and $\Delta=\sum_{i=1}^{N}\Delta_i$.
Thus we can obtain that $\dot{V}\le-lV+\Delta$, with   $l=\min\left\{2{c}_{1}^*,\cdots,2{c}_{N}^*,
\frac{2{\sigma}_{1}^*}{\lambda_{\max}\{\Gamma_1^{-1}\}},\cdots,
\frac{2{\sigma}_{N}^*}{\lambda_{\max}\{\Gamma_N^{-1}\}}\right\}$.

In what follows, we show that results i)-iii) in \emph{Theorem} 2 are ensured.
By following the similar lines as the proof of \emph{Theorem} 1, the results i)-ii) can be drawn. Next, we show that the result iii) is ensured.
Define $m_{k,l}^{i}(t)=x_{i,k}(t) -\bar{x}_{i,k}(t)$, $\forall t \in\left[t_{k,l}^{i},t_{k,l+1}^{i}\right)$, %$m_{k,l}^{j}(t)=x_{j,k}(t) -\bar{x}_{j,k}(t)$, $\forall t \in\left[t_{k,l}^{j},t_{k,l+1}^{j}\right)$, and $m_{u,l}^{i}(t)=v_i(t)-u_i(t)$, $\forall t \in\left[t_{u,l}^{i}, t_{u,l+1}^{i}\right)$, $i,j=1,\cdots,N\,(j\neq i)$, $k=1,\cdots,n_i$,
then it holds that
\begin{flalign}
&\frac{d\left|m_{k,l}^{i}\right|}{d t}
= \operatorname{sign}\left(m_{k,l}^{i}\right) \dot{m}_{k,l}^{i}\leq\left|\dot{m}_{k,l}^{i}\right|.
&\label{eq:48}
\end{flalign}
As $\bar{x}_{i,k}(t)$ remains unchanged for $t \in\left[t_{k,l}^{i}, t_{k,l+1}^{i}\right)$, one has $\left|\dot{m}_{k,l}^{i}\right|=\left|x_{i, k+1}{\rm{+}}\sum_{j=1}^{N} f_{i j, k}\right|,k=1,\cdots,n_i-1$ and  $\left|\dot{m}_{n_i,l}^{i}\right|=\left|u_{i}
{\rm{+}}\varphi_{i}^{T}\left(x_{i}\right){\theta_{i}}
{\rm{+}}\psi_i\left(x_{i}\right)+\sum_{j=1}^{N} f_{i j, n_{i}}\right|$.
Since $x_{i,k}$, $u_i$, $\varphi_{i}\left(x_{i}\right)$, $\psi_{i}\left(x_{i}\right)$ and $f_{ij,k}$, $k=1,\cdots,n_i$ are all bounded, it is derived that there exists positive constant $m_{0}^{i}$, such that $\left|\dot{m}_{k,l}^{i}\right| \leq m_{0}^{i}$, which implies that $t_{k,l+1}^{i}-t_{k,l}^{i}\geq \Delta x_{i,k}/m_{0}^{i}> T_{0}$. Similarly, it holds that $t_{k,l+1}^{j}-t_{k,l}^{j}> T_{1}$ and $t_{u,l+1}^{i}-t_{u,l}^{i}> T_2$, where $T_{0}$, $T_{1}$ and $T_{2}$ are positive constants. Therefore the Zeno behavior is excluded. $\hfill\blacksquare$ %The proof is completed.

${\textbf{Remark 6}}$.
%How the issue of non-differentiability of virtual control associated with backstepping design arising from intermittent full-state feedback is addressed, which in fact represents one of the key steps in control design and stability analysis.
To overcome the non-differentiability issue, we first develop a decentralized adaptive backstepping control scheme (\ref{eq:6})-(\ref{eq:9}) in a continuous fashion by properly treating the non-triangular structural uncertainties for the backstepping design, and simultaneously restrains the affects of the parameter-induced perturbation via freezing the time-varying parameters at the centers.
Afterwards, based upon the preceding scheme a decentralized adaptive event-triggered backstepping control scheme (\ref{eq:14})-(\ref{eq:17}) is proposed by replacing the states $x_{i,k}$ in the preceding scheme with $\bar{x}_{i,k}$, in which one key property utilized is that the partial derivatives $\xi_{k-1, l}^{i}\,(k=2,\cdots,n_i,l=1,\cdots,k)$ in each subsystem are all ensured to be constant. Finally, the crucial lemmas 1-2 are elaborately deduced with rigorous proofs for establishing stability condition under such replacement.

%Note that the continuous coordinate transformation  $z_{i,k}$ and virtual controllers $\alpha_{i,k-1}$ in (\ref{eq:4})-(\ref{eq:7}) are only used in the system stability analysis, in the sense that the final control $v_i$ and the parameter updating laws ${\dot{\hat{\theta}}}_{i}$ in (\ref{eq:16})-(\ref{eq:17}) utilize only the locally intermittent state signals $\bar{x}_{i,k}$, rather than $x_{i,k}$.

${\textbf{Remark 7}}$.
For handling the influence of the perturbation term $\varphi_{i}^T({x}_{i})\Delta_{\theta_i}{z}_{i,n_{i}}$  resulted from time-varying parameters, an alternative is the compensation approach adopted in \cite{chen2020adaptive}, which, however, is no longer applicable in this work. If a similar treatment is adopted, a compensation term $\frac{1}{2}\beta_{\theta_i} \varphi_{i}^{T}\left({\bar x}_{i}\right)\varphi_{i}\left({\bar x}_{i}\right){\bar z}_{i, n_{i}}$ will be added to the controller $v_i$, and thus the adverse effect term $\left(\varphi_{i}^{T}\left({\bar x}_{i}\right)\varphi_{i}\left({\bar x}_{i}\right){\bar z}_{i, n_{i}}^2-\varphi_{i}^{T}\left({x}_{i} \right)\varphi_{i}\left({x}_{i}\right){z}_{i, n_{i}}^2\right)$ caused by triggering error appears, which undoubtedly poses great difficulties in the stability analysis. To circumvent this problem, we handle such term in a non-compensatory manner with the aid of \emph{Lemma} 1, as shown in (\ref{eq:44}), from which we obtain that the upper bound of $\left|\varphi_{i}^T({x}_{i})\Delta_{\theta_i}{z}_{i,n_{i}}\right|$ relies on $\left\|z_{i}\right\|^2$ only, thereby can be incorporated into the negative term $-\sum_{j=1}^{N}{c}_{j}^*\left\|z_{j}\right\|^{2}$ in (\ref{eq:47}).

${\textbf{Remark 8}}$.
Compared with existing results for time-varying systems \cite{goel2022composite,arefi2010adaptive}, where the adverse effects induced by time-varying parameters are directly addressed, this work obviously exhibits its merit because in accommodating the impact of the time-varying parameters, the somewhat restrictive conditions, such as the initial excitation \cite{goel2022composite}, or the matched uncertainties \cite{arefi2010adaptive}, are completely removed. In addition, the existing methods related to event-triggered adaptive control \cite{YAO2021527,yao2021event}, although based upon non-triangular systems, are inapplicable to the setting of this work that is more comprehensive (i.e., both state and input are triggered) yet involves time-varying parameters.
Moreover, the limitation of the triangular condition as typically imposed in current related works \cite{zhang2021adaptive,sun2022Distributed,DEPERSIS20132116} is eliminated in this work, substantially  broadening its scope of applications.

%${\textbf{Remark 9}}$.
%%Note that the proposed modified congelation of variables based approach not only requires less restrictive assumptions as compared to the related works  \cite{goel2022composite,arefi2010adaptive}, but is also compatible with most adaptive control schemes involving parameter estimation, in the sense that it does not change the original parameter update law designed for time-invariant systems.
%Note that the solution in this work only obtains a globally uniformly ultimately bounded (GUUB) result of the closed-loop system (instead of attaining the globally asymptotical stability), which is a consequence of the involvement of the non-triangular structural uncertainties and the event-triggering mechanism.
%As the focus of this work is primarily on developing a decentralized event-triggered adaptive backstepping control for nonlinear non-triangular interconnected systems with time-varying parameters, it opens the door for further improvements on system performance in the future study.

\section{Simulation Verification}
To verify the efficiency of the proposed control method, we consider the following interconnected system with two subsystems: %\cite{Caijianping}
\begin{flalign}
\dot{x}_{i, 1}=\,&x_{i, 2}+\sum_{j=1}^{2} f_{i j, 1}&\nonumber\\
\dot{x}_{i, 2}=\,&u_{i}+\varphi_{i}\left(x_i\right) \theta_{i}(t)+\sum_{j=1}^{2} f_{i j, 2},\,\,y_{1}=x_{i, 1}& \label{eq:50}
\end{flalign}
for $i=1,2$. In the simulation, we set the initial states $x_{1,1}(0)=0.2$, $x_{1,2}(0)=0.2$, $x_{2,1}(0)=0.1$, $x_{2,2}(0)=0.1$, the design parameters $c_{1,1}=0.5$, $c_{1,2}=0.3$, $c_{2,1}=1.8$, $c_{2,2}=1.5$, $\sigma_{1}=0.001$, $\sigma_{2}=0.001$,
$\Gamma_1=0.5$, $\Gamma_2=0.5$, $\varpi_{ii,k,l}=1\,(i,k,l=1,2)$, $\varpi_{ij,k,l}=1\,(i,j,k,l=1,2)$ for $i\neq j$, the time-varying parameters $\theta_{1}(t)=0.1+0.1\sin(0.2t)$, $\theta_{2}(t)=0.1+0.1\cos(0.2t)$, $\varphi_{1}=0.2\left(x_{1,1}^{2}+x_{1,2}\right)+3 \cos \left(x_{1,1} x_{1,2}\right)$, $\varphi_{2}=0.2\left(x_{2,1}^{2}+x_{2,2}\right)+3 \cos \left(x_{2,1} x_{2,2}\right)$, the nonlinear
functions $f_{11,1}=0.1\sin (u_1u_2)\sqrt{x_{1,1}^{2}+x_{1,2}^{2}}$, $f_{12,1}=0.15 \sqrt{x_{2,1}^{2}+x_{2,2}^{2}}$,
$f_{11,2}=0.1 \sqrt{x_{1,1}^{2}+x_{1,2}^{2}}$, $f_{12,2}=0.15 \sin
\left(\sqrt{x_{2,1}^{2}+x_{2,2}^{2}}\right)$, $f_{21,1}=0.15 \sqrt{x_{1,1}^{2}+x_{1,2}^{2}}$, $f_{22,1}=0.1 \cos (u_1u_2)
\sqrt{x_{2,1}^{2}+x_{2,2}^{2}}$,
$f_{21,2}=0.15 \sqrt{x_{1,1}^{2}+x_{1,2}^{2}}$, $f_{22,2}=$ $0.1 \ln \left(1+\sqrt{x_{2,1}^{2}+x_{2,2}^{2}}\right)$, which do not meet the triangular structure requirements. From the given $f_{i j, k}$ in (44), we set $\hbar_{i j, k}$ and $\epsilon_{i j, k}=0$.
In addition, to test the effect of triggering thresholds on the tracking performance, we set the triggering thresholds as: 1) $\Delta x_{1,1}=0.001$, $\Delta x_{1,2}=0.002$, $\Delta x_{2,1}=0.002$, $\Delta x_{2,2}=0.002$, $\Delta {u_1}=0.01$, $\Delta {u_2}=0.01$; 2) $\Delta x_{1,1}'=0.005$, $\Delta x_{1,2}'=0.005$, $\Delta x_{2,1}'=0.003$, $\Delta x_{2,2}'=0.003$, $\Delta u_1'=0.03$, $\Delta u_2'=0.03$, and the same set of other design parameters are used.

The results are presented in Fig. 2. Fig. 2 (a)-(b) show the state trajectories $x_{1,k}$ and $x_{2,k}\,(k=1,2)$, respectively. Fig. 2 (c) gives the control input $u_i$. Fig. 2 (d) shows the time-varying adaptive estimated parameter vector $\hat{\theta}_i(t)$. Fig. 2 (e)-(f) gives state trajectories $x_{1,k}$ and $x_{2,k}$ for the case of increasing triggering thresholds, respectively.
The triggered times of $x_{i,k}\,(i,k=1,2)$ and $u_i$ for different triggering thresholds are presented in Fig. 2 (g)-(h).
%The number of triggering events of $u_1$, $u_2$, $x_{1,1}$, $x_{1,2}$, $x_{2,1}$ and  $x_{2,2}$ are: Case 1) 234, 306, 221, 291, 58, 231; and Case 2) 103, 221, 47, 149, 40, 181, respectively.
From the simulation results, it can be concluded that all signals in the closed-loop systems are globally uniformly bounded, meanwhile all the subsystem outputs are steered into a residual set around zero. Besides, it holds that the larger the triggering thresholds, the smaller the triggering times. Nevertheless, it can be observed that the system performance is degraded to some extent.

%\begin{table}[!htbp]%{Table 1}
%\centering
%\caption{The number of triggering events}\label{tab:aStrangeTable}
%\label{tab:1}
%\begin{tabular}{cccc}
%\hline\noalign{\smallskip}
%\multicolumn{2}{l}{\textbf{\emph{\qquad\qquad \qquad \qquad Case 1 \,   \qquad  \qquad\qquad \, Case 2 \, }}}\\
%\noalign{\smallskip}\hline\noalign{\smallskip}
%$u_1$ $\:$ &   \qquad\qquad $\:234$ \qquad  \qquad \qquad \qquad \,\,\,$\:103$ \\
%\noalign{\smallskip}
%$u_2$ $\:$ &  \qquad\qquad $\:306$ \qquad  \qquad \qquad  \qquad \,\,\,$\:221$\\
%\noalign{\smallskip}
%$x_{1,1}$ $\:$ &  \qquad\quad\,\, $\:221$ \qquad  \qquad \qquad \qquad \,\,\,\,$\:47$\\
%\noalign{\smallskip}
%$x_{1,2}$ $\:$ &  \qquad\qquad $\:291$ \qquad   \qquad \qquad \qquad\, $\:149$\\
%\noalign{\smallskip}
%$x_{2,1}$ $\:$ &  \qquad\quad \,\,\, $\:58$ \qquad  \qquad \qquad \qquad\,\,\,\, $\:40$\\
%\noalign{\smallskip}
%$x_{2,2}$ $\:$&  \qquad\quad \,\,\,\, $\:231$ \qquad  \qquad \qquad \quad \,\,\,\,\,\,\,\,\,$\:181$\\
%\noalign{\smallskip}\hline
%\end{tabular}
%\end{table}

\begin{figure*}[t]
\centering{
\subfigure[States $x_{1,1}$ and $x_{1,2}$.]
{\includegraphics[width=1.75in]{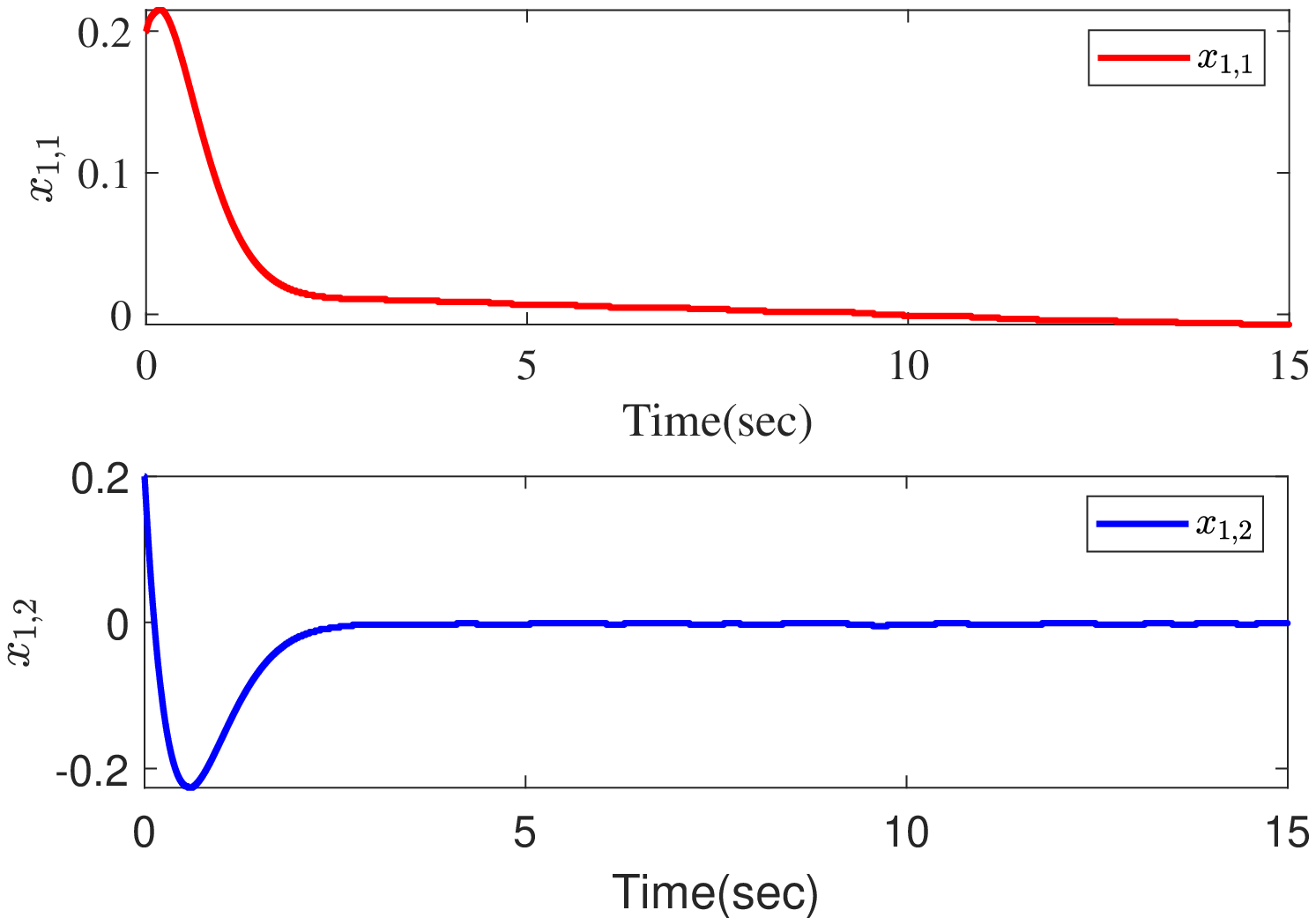}}
\subfigure[States $x_{2,1}$ and $x_{2,2}$.]
{\includegraphics[width=1.75in]{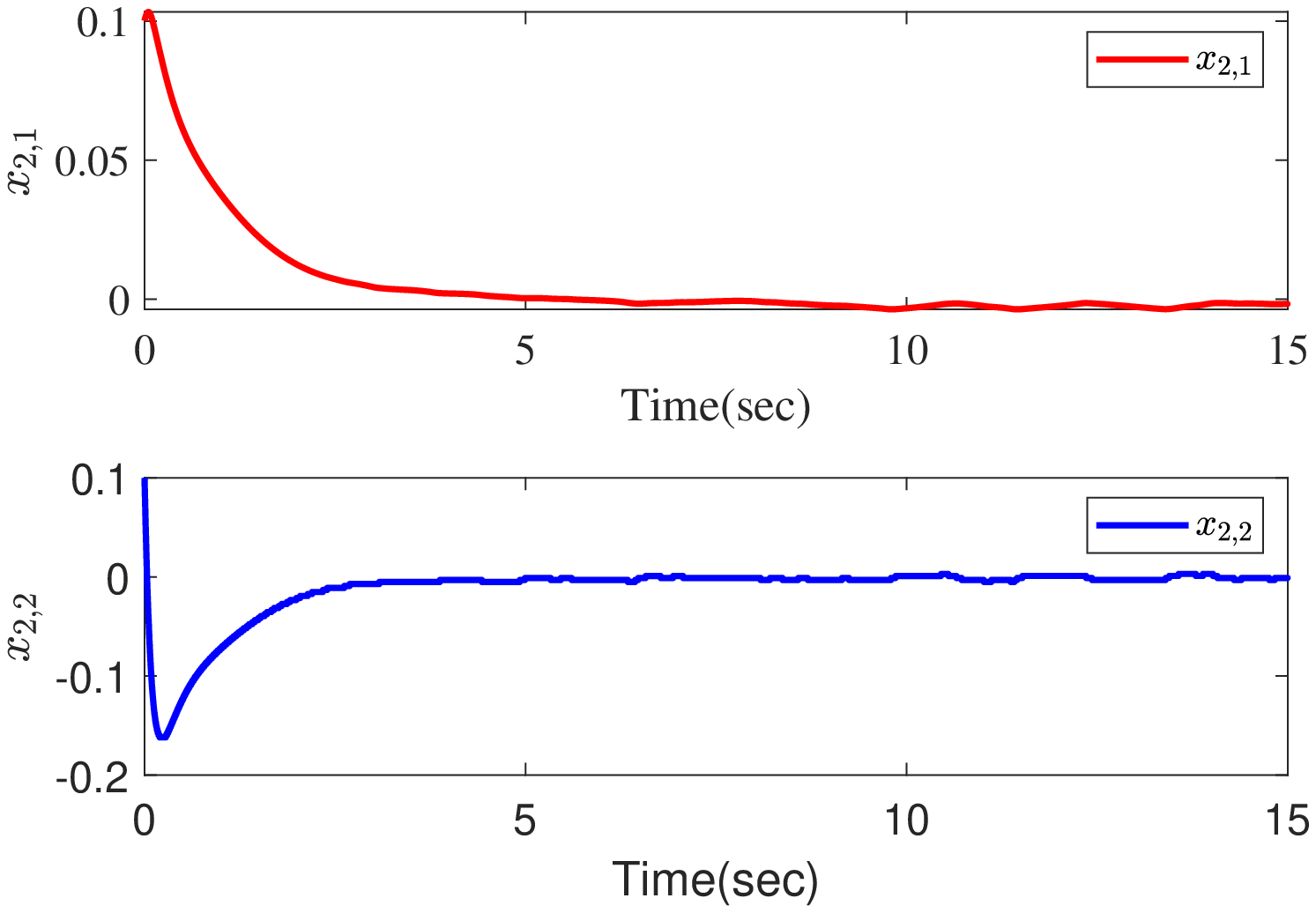}}
\subfigure[Control input $u_i$.]
{\includegraphics[width=1.75in]{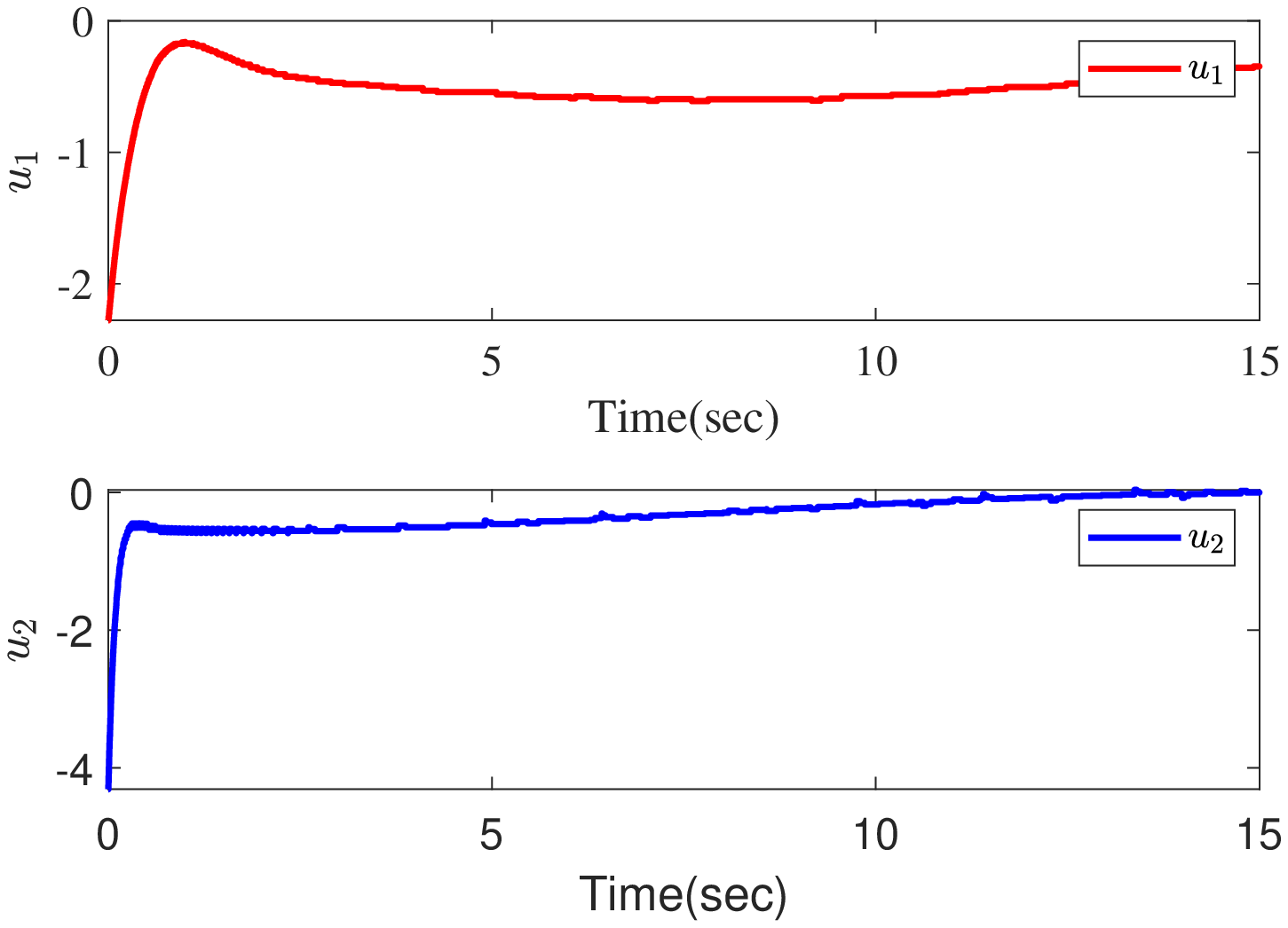}}
\subfigure[Time-varying adaptive estimated parameter vector $\hat{\theta}_i(t)$.]
{\includegraphics[width=1.75in]{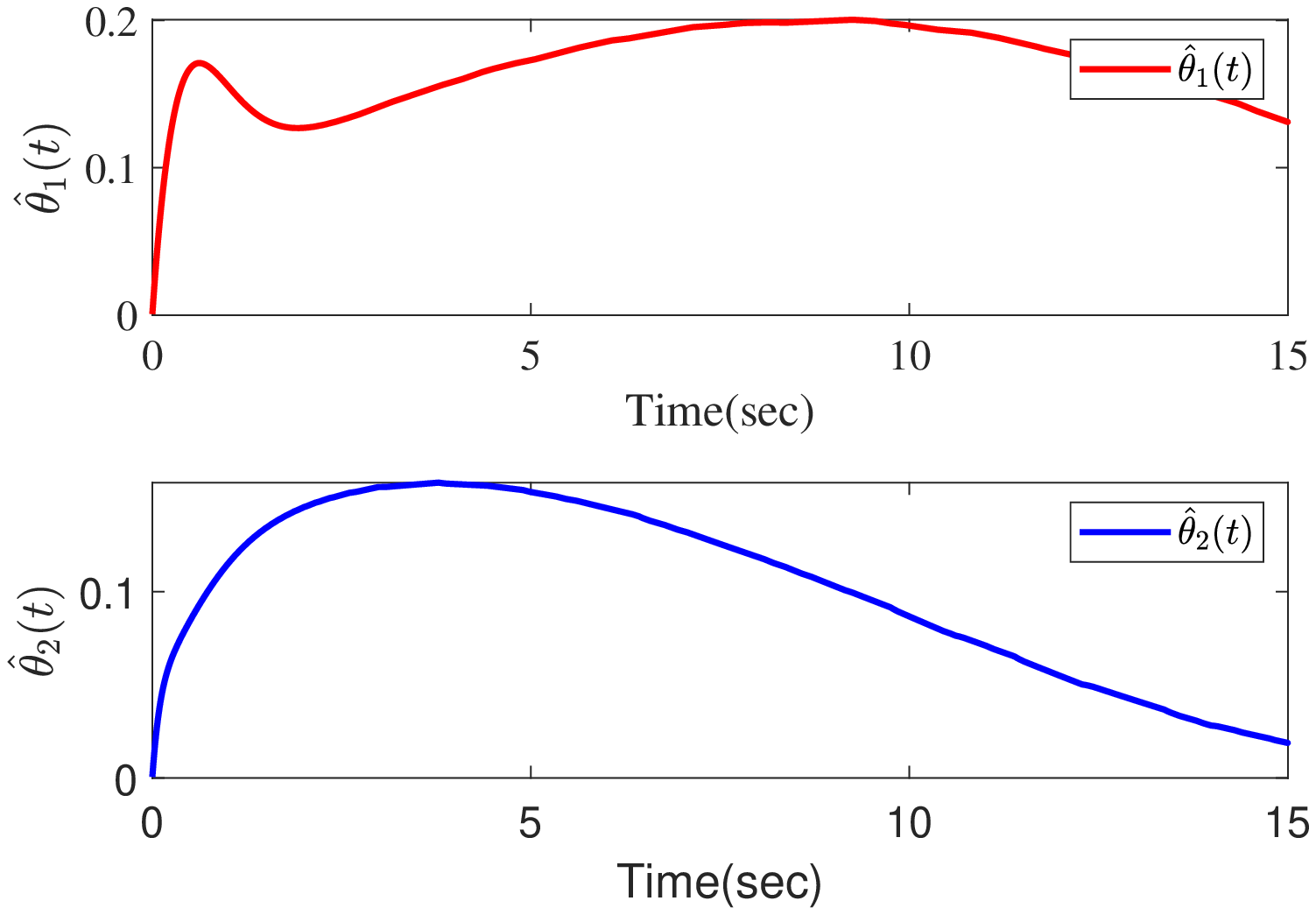}}
\subfigure[$x_{1,1}$ and $x_{1,2}$ for the case of increasing triggering thresholds.]
{\includegraphics[width=1.75in]{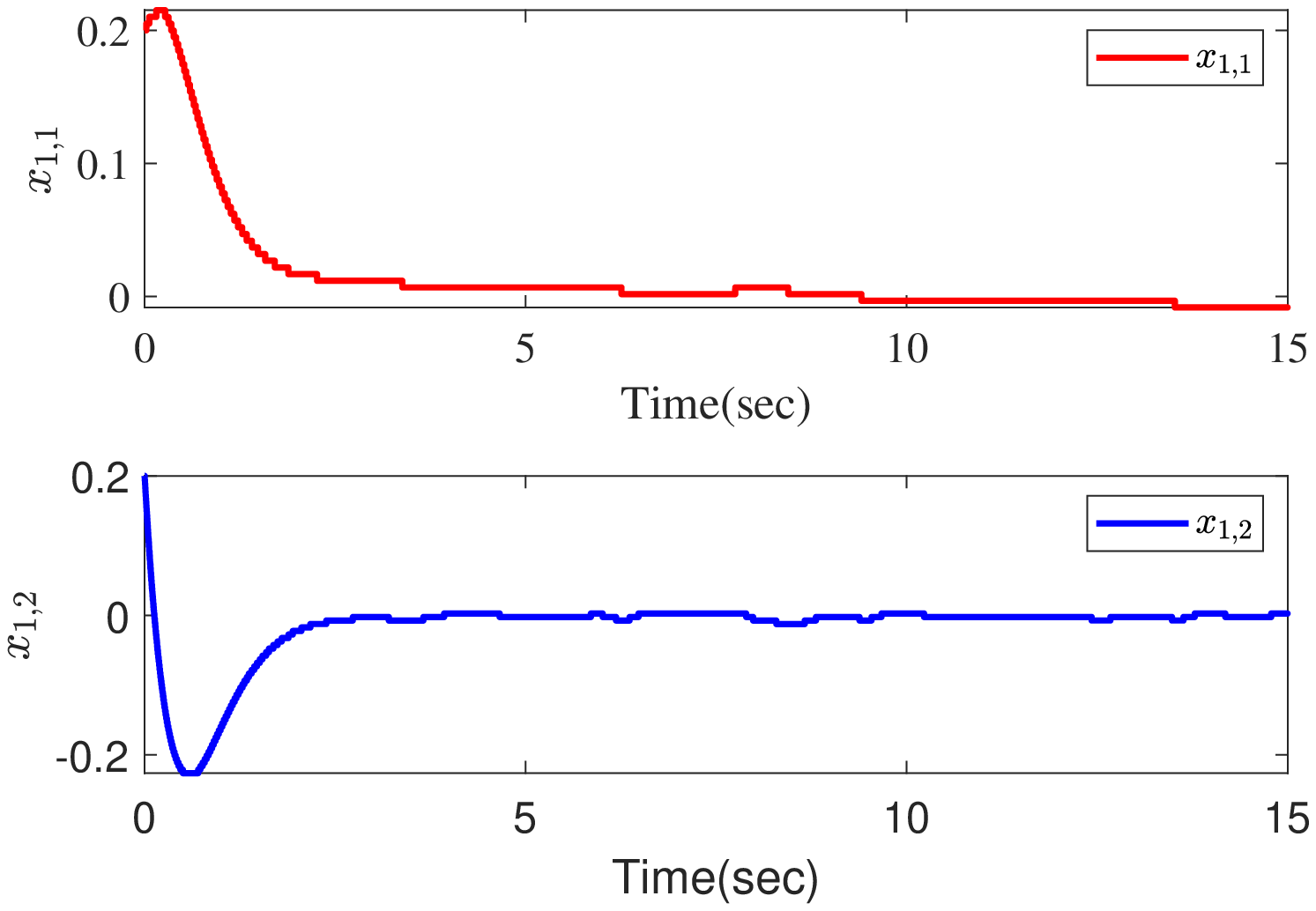}}
\subfigure[$x_{2,1}$ and $x_{2,2}$ for the case of increasing triggering thresholds.]
{\includegraphics[width=1.75in]{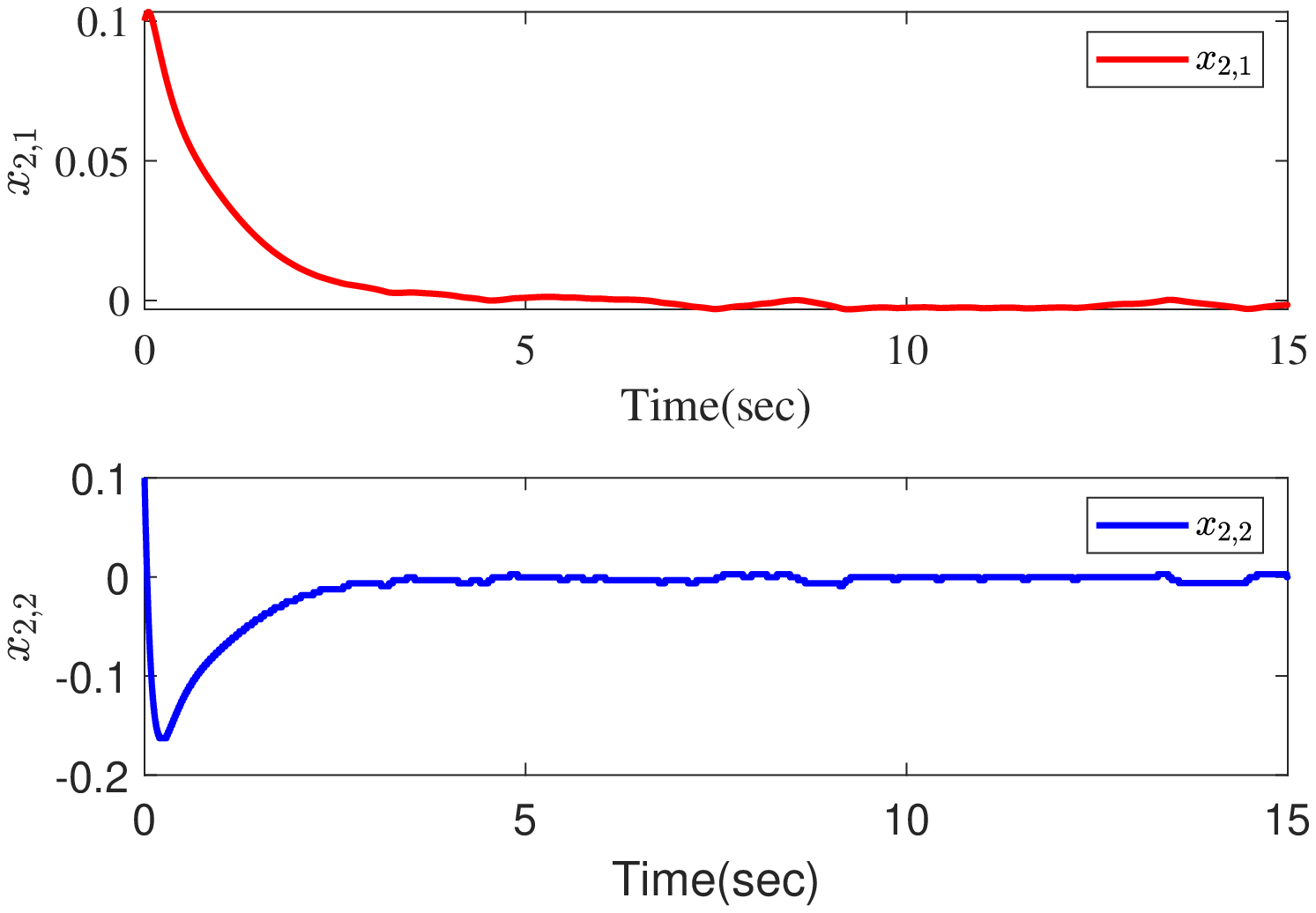}}
%\subfigure[$u_i$ for the case of increasing triggering thresholds.]
%{\includegraphics[width=1.7in]{figure_20220603/u1u2_1.eps}}
\subfigure[Triggering times of ${x}_{i,k}\,(i,k=1,2)$ for different triggering thresholds.]
{\includegraphics[width=1.75in]{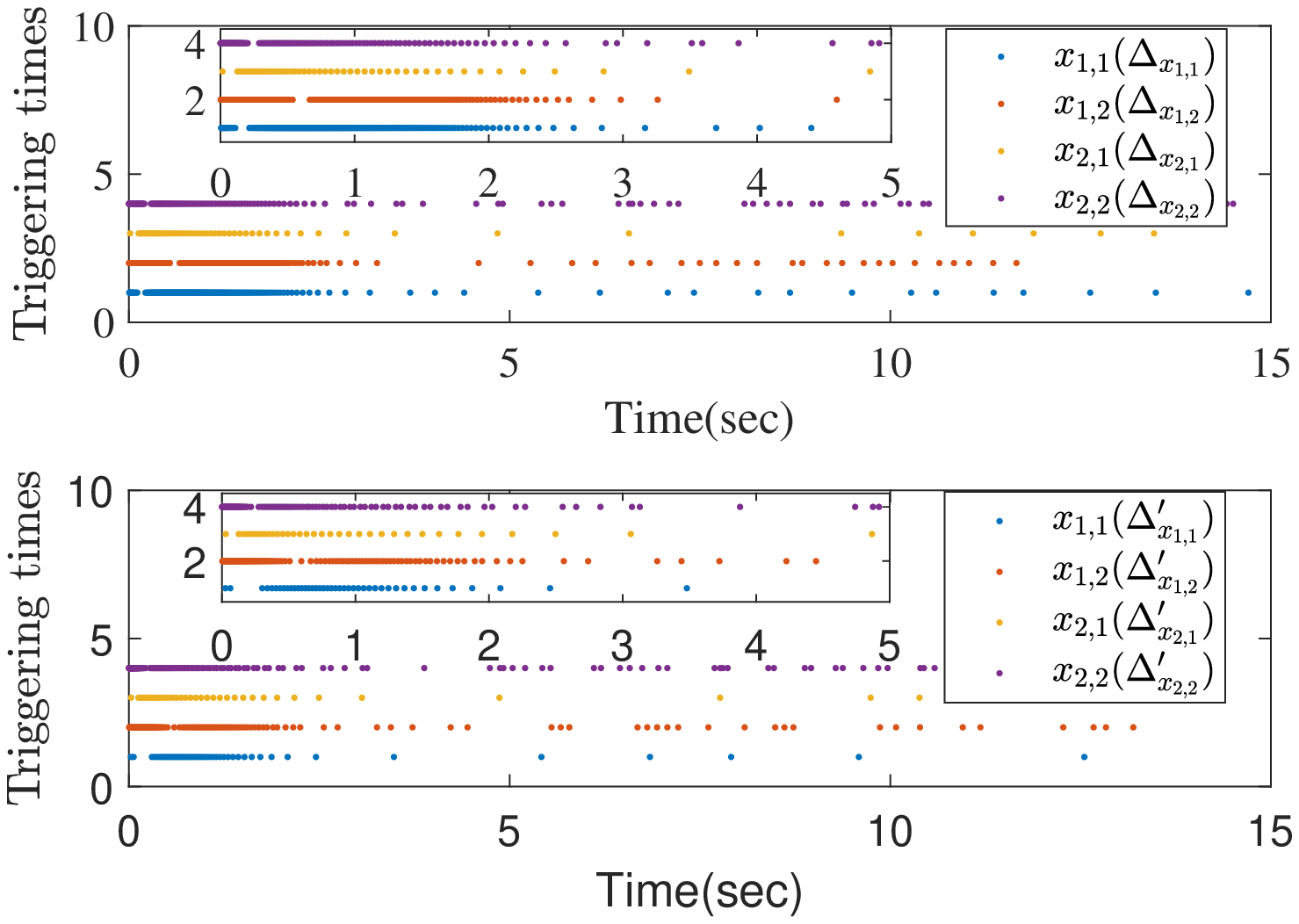}}
\subfigure[Triggering times of ${u}_{i}$ for different triggering thresholds.]
{\includegraphics[width=1.75in]{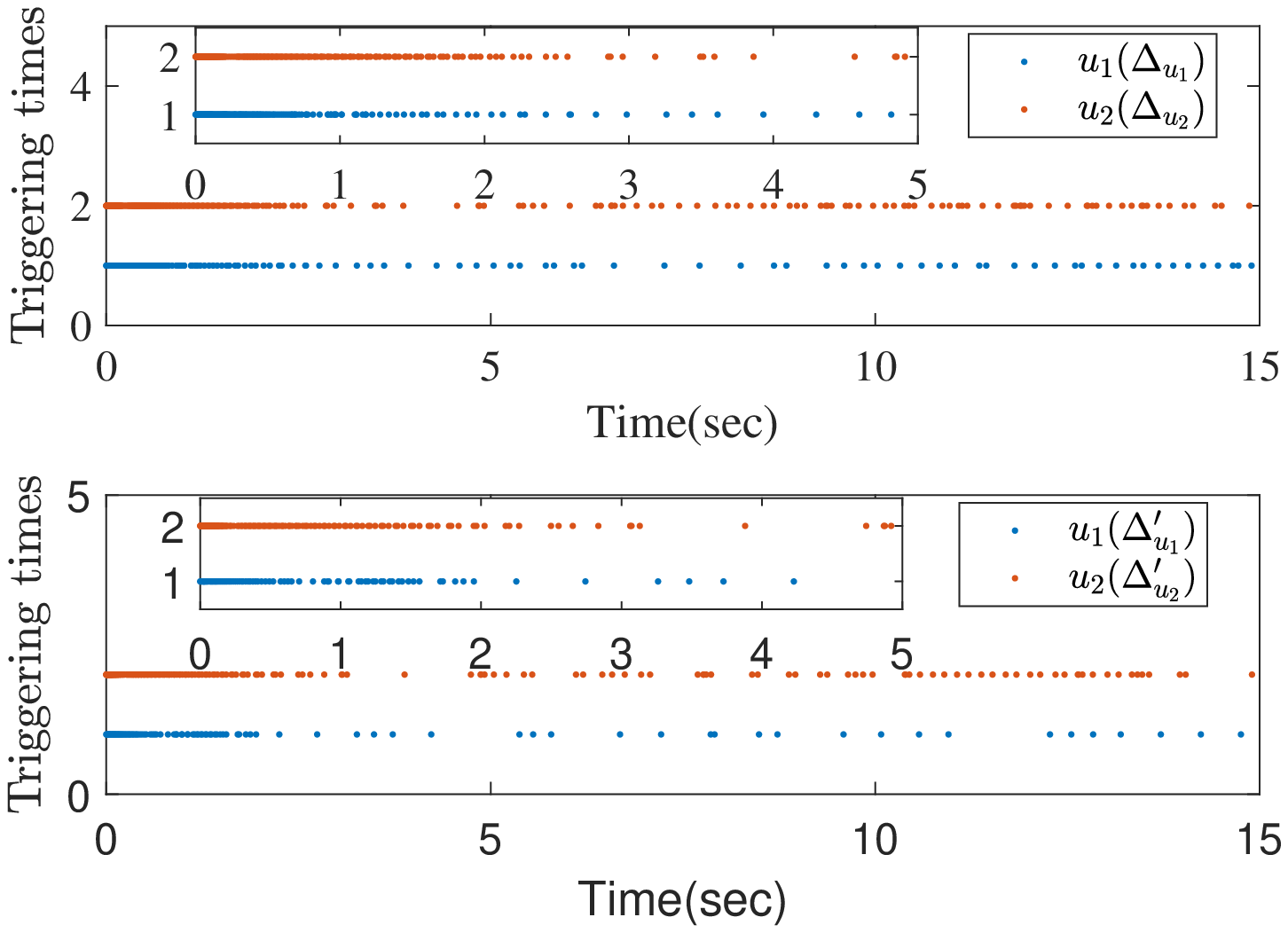}}}
\caption{Simulation results by using the proposed event-triggered control scheme.}
 \label{simulation_results}
\end{figure*}

\section{Conclusion}
This work presents a decentralized adaptive backstepping control scheme for non-triangular nonlinear time-varying systems via intermittent state feedback. The major technical challenge in developing such control strategy is to obviate the non-differentiability of the virtual control arising from intermittent state feedback, while coping with the non-triangular structural uncertainties and unknown time-varying parameters. By using the results established in the lemmas with rigorous proofs, it is shown that the closed-loop signal is globally uniformly bounded without Zeno behavior, and at the same time all the subsystem outputs are steered into an assignable residual set around zero. An interesting topic for future research is the consideration of the tracking control problem for such system.

\begin{appendices}
\section{}
\textbf{Proof of Lemma 1}.
From (\ref{eq:4})-(\ref{eq:7}), it is seen that
\begin{flalign}
{z_{i,1}} =\,& {x_{i,1}}& \label{eq:62}\\
{z_{i,2}} =\,& {x_{i,2}}+c_{i, 1} z_{i, 1}+\frac{1}{4 \varpi_{i i, 1,1}} z_{i, 1}+\frac{1}{4 \varpi_{i i, 1,2}} z_{i, 1} &\nonumber\\
&+\sum_{j \neq i} \frac{1}{4 \varpi_{i j, 1,1}} z_{i, 1}+\sum_{j \neq i}\frac{1}{4 \varpi_{i j, 1,2}} z_{i, 1}& \label{eq:63}\\
{z_{i,k}} =\,& {x_{i,k}} +c_{i, k} z_{i, k}+\sum_{l=1}^{k-1} \xi_{k-1, l}^{i} x_{i, l+1}+z_{i, k-1} &\nonumber\\
&+\sum_{j=1}^{N}\sum_{l=1}^{k-1}
\left(\frac{\left(\xi_{k-1, l}^{i}\right)^{2}}{4 \varpi_{i j, k, 1}}+\frac{\left(\xi_{k-1, l}^{i}\right)^{2}}{4 \varpi_{i j, k, 2}}\right)z_{i, k}&\nonumber\\
&+\sum_{j=1}^{N}\left(\frac{1}{4 \varpi_{i j, k, 1}}+\frac{1}{4 \varpi_{i j, k,2}}\right)z_{i, k},k = 3, \cdots ,n_i& \label{eq:64}
\end{flalign}
Then it can derived that
\begin{flalign}
&A_{i}(c_{i, \tau}, \varpi_{i j, \tau, 1}, \varpi_{i j, \tau, 2}) x_{i}=B_{i}(c_{i, \tau}, \varpi_{i j, \tau, 1}, \varpi_{i j, \tau, 2}) z_{i}
& \label{eq:65}
\end{flalign}
with
\begin{flalign}
&A_{i}=\left(\begin{array}{ccccc}
1 & 0 & 0 & \cdots & 0 \\
0 & 1 & 0 & \cdots & 0 \\
\vdots & \vdots & \vdots & & \vdots \\
0 & -\xi_{n_{i}-2,2}^{i} & -\xi_{n_{i}-2,3}^{i} & \cdots & 1
\end{array}\right) & \label{eq:66}
\end{flalign}
\begin{flalign}
&B_{i}=\left(\begin{array}{ccccccc}
1 & 0 & 0 & \cdots & 0 & 0 & 0 \\
B_{i,1} & 1 & 0 & \cdots & 0 & 0 & 0 \\
\vdots & \vdots & \vdots & & \vdots & \vdots & \vdots \\
0 & 0 & 0 & \cdots & -1 & B_{i,2} & 1
\end{array}\right). & \label{eq:67}
\end{flalign}
which are constant matrices, and their components are the functions of parameters $c_{i,\tau}$, $\varpi_{i j, \tau, 1}$ and $\varpi_{i j, \tau, 2}$, $\tau=1, \cdots, n_{i}$,  where $B_{i,1}=-c_{i, 1}-\sum_{j=1}^{N}\left(\frac{1}{4 \varpi_{i j, 1,1}}+\frac{1}{4 \varpi_{i j, 1,2}}\right)$ and $B_{i,2}=-c_{i, n_{i}-1}-\sum_{j=1}^{N} \sum_{l=1}^{n_{i}-2}\left(\frac{\left(\xi_{n_{i}-2, l}^{i}\right)^{2}}{4 \varpi_{i j, n_{i}-1,1}}+\frac{\left(\xi_{n_{i}-2, l}^{i}\right)^{2}}{4 \varpi_{i j, n_{i}-1,2}}\right)-\sum_{j=1}^{N}\left(\frac{1}{4 \varpi_{i j, n_{i}-1,1}}+\frac{1}{4 \varpi_{i j, n_{i}-1,2}}\right)$. Clearly $A_{i}$ is an invertible matrix, which implies that $x_{i}=A_{i}^{-1} B_{i} z_{i}$, then it holds that $\left\|x_{i}\right\| \leq\left\|A_{i}^{-1}B_{i}\right\|_{F}\left\|z_{i}\right\|$. $\hfill{\blacksquare}$
%This completes the proof.

\section{}
\textbf{Proof of Lemma 2}.
From (\ref{eq:4}) and (\ref{eq:12}), it is readily seen that
\begin{flalign}
\left| {{z_{i,1}} - {{\bar z}_{i,1}}} \right| =& \left| {{x_{i,1}} - {{\bar x}_{i,1}}} \right| \le \Delta {x_{i,1}}\buildrel \Delta \over = \Delta{z_{i,1}}.& \label{eq:70}
\end{flalign}
According to (\ref{eq:6}), (\ref{eq:14}) and (\ref{eq:70}),  it follows that
\begin{flalign}
\left|\alpha_{i, 1}- \bar{\alpha}_{i, 1}\right|\le &\left(c_{i, 1}+\frac{1}{4 \varpi_{i i, 1,1}}+\frac{1}{4 \varpi_{i i, 1,2}}+\sum_{j \neq i}\frac{1}{4 \varpi_{i j, 1,1}}\right.&\nonumber\\
&\left.+\sum_{j \neq i}\frac{1}{4 \varpi_{i j, 1,2}} \right)\Delta z_{i, 1} \buildrel \Delta \over = \Delta{\alpha_{i,1}}.
& \label{eq:71}
\end{flalign}
Similarly, from (\ref{eq:5}), (\ref{eq:7}),  (\ref{eq:13}) and (\ref{eq:15}), it holds that
\begin{flalign}
\left| {{z_{i,k}} - {{\bar z}_{i,k}}} \right| %\le& \left| {{x_{i,k}} - {{\bar x}_{i,k}}} \right| + \left| {{\alpha _{i,{k-1}}} - {{\bar \alpha }_{i,{k-1}}}} \right|& \nonumber\\
\le &\Delta {x_{i,k}} + \Delta {\alpha_{i,{k-1}}}\buildrel \Delta \over=\Delta{z_{i,k}}&\label{eq:72}\\
\left|\alpha_{i, k}{\rm{-}} \bar{\alpha}_{i, k}\right|\le & \,c_{i, k} \Delta z_{i, k}+\sum_{l=1}^{k-1} \left|\xi_{k-1, l}^{i}\right| \Delta x_{i, l+1}+\Delta z_{i, k-1}&\nonumber \\
%&\,+ \left(\sum_{j=1}^{N}
%\left(\frac{1}{4 \varpi_{i j, k, 1}}{\rm{+}}\frac{1}{4 \varpi_{i j, k,2}}\right)+\sum_{j=1}^{N}
%\sum_{l=1}^{k-1}\right.&\nonumber \\
%&\left.\left(\frac{\left(\xi_{k-1, l}^{i}\right)^{2}}{4 \varpi_{i j, k, 1}}{\rm{+}}\frac{\left(\xi_{k-1, l}^{i}\right)^{2}}{4 \varpi_{i j, k, 2}}\right)\right)\Delta z_{i, k}{\rm{\buildrel\Delta \over=}} \Delta{\alpha_{i,k}}&\nonumber \\
%
&\,+ \Delta z_{i, k}\left(\sum_{j=1}^{N}
\sum_{l=1}^{k-1}\left(\frac{\left(\xi_{k-1, l}^{i}\right)^{2}}{4 \varpi_{i j, k, 1}}+\frac{\left(\xi_{k-1, l}^{i}\right)^{2}}{4 \varpi_{i j, k, 2}}\right)\right.&\nonumber \\
&\left. +\sum_{j=1}^{N}
\left(\frac{1}{4 \varpi_{i j, k, 1}}+\frac{1}{4 \varpi_{i j, k,2}}\right) \right)\buildrel\Delta \over= \Delta{\alpha_{i,k}}
& \label{eq:73}
\end{flalign}
for $k=2,\cdots,n_i$. %This completes the proof.
$\hfill{\blacksquare}$
\end{appendices}

\bibliographystyle{IEEEtran}
\bibliography{interconnected_systems_20220412}

% Generated by IEEEtran.bst, version: 1.13 (2008/09/30)
\begin{thebibliography}{10}
\providecommand{\url}[1]{#1}
\csname url@samestyle\endcsname
\providecommand{\newblock}{\relax}
\providecommand{\bibinfo}[2]{#2}
\providecommand{\BIBentrySTDinterwordspacing}{\spaceskip=0pt\relax}
\providecommand{\BIBentryALTinterwordstretchfactor}{4}
\providecommand{\BIBentryALTinterwordspacing}{\spaceskip=\fontdimen2\font plus
\BIBentryALTinterwordstretchfactor\fontdimen3\font minus
  \fontdimen4\font\relax}
\providecommand{\BIBforeignlanguage}[2]{{%
\expandafter\ifx\csname l@#1\endcsname\relax
\typeout{** WARNING: IEEEtran.bst: No hyphenation pattern has been}%
\typeout{** loaded for the language `#1'. Using the pattern for}%
\typeout{** the default language instead.}%
\else
\language=\csname l@#1\endcsname
\fi
#2}}
\providecommand{\BIBdecl}{\relax}
\BIBdecl

\bibitem{ioannou1986decentralized}
P.~Ioannou, ``Decentralized adaptive control of interconnected systems,''
  \emph{IEEE Transactions on Automatic Control}, vol.~31, no.~4, pp. 291--298,
  1986.

\bibitem{harmand2005optimal}
J.~Harmand and D.~Dochain, ``The optimal design of two interconnected (bio)
  chemical reactors revisited,'' \emph{Computers \& chemical engineering},
  vol.~30, no.~1, pp. 70--82, 2005.

\bibitem{wen1992global}
C.~Wen and D.~J. Hill, ``Global boundedness of discrete-time adaptive control
  just using estimator projection,'' \emph{Automatica}, vol.~28, no.~6, pp.
  1143--1157, 1992.

\bibitem{jiang2000decentralized}
Z.-P. Jiang, ``Decentralized and adaptive nonlinear tracking of large-scale
  systems via output feedback,'' \emph{IEEE Transactions on Automatic control},
  vol.~45, no.~11, pp. 2122--2128, 2000.

\bibitem{4118474}
P.~Antsaklis and J.~Baillieul, ``Special issue on technology of networked
  control systems,'' \emph{Proceedings of the IEEE}, vol.~95, no.~1, pp. 5--8,
  2007.

\bibitem{heemels2012periodic}
W.~H. Heemels, M.~Donkers, and A.~R. Teel, ``Periodic event-triggered control
  for linear systems,'' \emph{IEEE Transactions on automatic control}, vol.~58,
  no.~4, pp. 847--861, 2012.

\bibitem{astrom2008event}
K.~J. Astr{\"o}m, ``Event based control,'' in \emph{Analysis and design of
  nonlinear control systems}.\hskip 1em plus 0.5em minus 0.4em\relax Springer,
  2008, pp. 127--147.

\bibitem{SEURET201647}
A.~Seuret, C.~Prieur, S.~Tarbouriech, and L.~Zaccarian, ``Lq-based
  event-triggered controller co-design for saturated linear systems,''
  \emph{Automatica}, vol.~74, pp. 47--54, 2016.

\bibitem{zhu2014event}
W.~Zhu, Z.-P. Jiang, and G.~Feng, ``Event-based consensus of multi-agent
  systems with general linear models,'' \emph{Automatica}, vol.~50, no.~2, pp.
  552--558, 2014.

\bibitem{tabuada2007event}
P.~Tabuada, ``Event-triggered real-time scheduling of stabilizing control
  tasks,'' \emph{IEEE Transactions on Automatic Control}, vol.~52, no.~9, pp.
  1680--1685, 2007.

\bibitem{adaldo2015event}
A.~Adaldo, F.~Alderisio, D.~Liuzza, G.~Shi, D.~V. Dimarogonas, M.~Di~Bernardo,
  and K.~H. Johansson, ``Event-triggered pinning control of switching
  networks,'' \emph{IEEE Transactions on Control of Network Systems}, vol.~2,
  no.~2, pp. 204--213, 2015.

\bibitem{xing2016event}
L.~Xing, C.~Wen, Z.~Liu, H.~Su, and J.~Cai, ``Event-triggered adaptive control
  for a class of uncertain nonlinear systems,'' \emph{IEEE transactions on
  automatic control}, vol.~62, no.~4, pp. 2071--2076, 2016.

\bibitem{ghodrat2018local}
M.~Ghodrat and H.~J. Marquez, ``On the local input--output stability of
  event-triggered control systems,'' \emph{IEEE Transactions on Automatic
  Control}, vol.~64, no.~1, pp. 174--189, 2018.

\bibitem{ABDELRAHIM201796}
M.~Abdelrahim, R.~Postoyan, J.~Daafouz, and D.~Nešić, ``Robust
  event-triggered output feedback controllers for nonlinear systems,''
  \emph{Automatica}, vol.~75, pp. 96--108, 2017.

\bibitem{ZHANG2022110283}
Z.~Zhang, C.~Wen, K.~Zhao, and Y.~Song, ``Decentralized adaptive control of
  uncertain interconnected systems with triggering state signals,''
  \emph{Automatica}, vol. 141, p. 110283, 2022.

\bibitem{wang2020adaptive}
W.~Wang, C.~Wen, J.~Huang, and J.~Zhou, ``Adaptive consensus of uncertain
  nonlinear systems with event triggered communication and intermittent
  actuator faults,'' \emph{Automatica}, vol. 111, p. 108667, 2020.

\bibitem{wang2021adaptive}
W.~Wang, J.~Long, J.~Zhou, J.~Huang, and C.~Wen, ``Adaptive backstepping based
  consensus tracking of uncertain nonlinear systems with event-triggered
  communication,'' \emph{Automatica}, vol. 133, p. 109841, 2021.

\bibitem{zhang2021adaptive}
Z.~Zhang, C.~Wen, L.~Xing, and Y.~Song, ``Adaptive event-triggered control of
  uncertain nonlinear systems using intermittent output only,'' \emph{IEEE
  Transactions on Automatic Control}, 2021.

\bibitem{sun2022Distributed}
L.~Sun, X.~Huang, and Y.~Song, ``Distributed event-triggered control of
  networked strict-feedback systems via intermittent state feedback,''
  \emph{IEEE Transactions on Automatic Control, Review}, 2022.

\bibitem{DEPERSIS20132116}
C.~{De Persis}, R.~Sailer, and F.~Wirth, ``Parsimonious event-triggered
  distributed control: A zeno free approach,'' \emph{Automatica}, vol.~49,
  no.~7, pp. 2116--2124, 2013.

\bibitem{zhang2003backstepping}
Y.~Zhang, B.~Fidan, and P.~A. Ioannou, ``Backstepping control of linear
  time-varying systems with known and unknown parameters,'' \emph{IEEE
  Transactions on Automatic Control}, vol.~48, no.~11, pp. 1908--1925, 2003.

\bibitem{goel2022composite}
R.~Goel and S.~B. Roy, ``Composite adaptive control for time-varying systems
  with dual adaptation,'' \emph{arXiv preprint arXiv:2206.01700}, 2022.

\bibitem{arefi2010adaptive}
M.~Arefi and M.~Jahed-Motlagh, ``Adaptive robust synchronization of rossler
  systems in the presence of unknown matched time-varying parameters,''
  \emph{Communications in Nonlinear Science and Numerical Simulation}, vol.~15,
  no.~12, pp. 4149--4157, 2010.

\bibitem{cai2022decentralized}
J.~Cai, C.~Wen, L.~Xing, and Q.~Yan, ``Decentralized backstepping control for
  interconnected systems with non-triangular structural uncertainties,''
  \emph{IEEE Transactions on Automatic Control}, 2022.

\bibitem{zhou2022event}
Y.~Zhou, D.~Li, Y.~Xi, and F.~Gao, ``Event-triggered distributed robust model
  predictive control for a class of nonlinear interconnected systems,''
  \emph{Automatica}, vol. 136, p. 110039, 2022.

\bibitem{pagilla2006decentralized}
P.~R. Pagilla, N.~B. Siraskar, and R.~V. Dwivedula, ``Decentralized control of
  web processing lines,'' \emph{IEEE Transactions on control systems
  technology}, vol.~15, no.~1, pp. 106--117, 2006.

\bibitem{cai2016adaptive}
J.~Cai, C.~Wen, H.~Su, Z.~Liu, and L.~Xing, ``Adaptive backstepping control for
  a class of nonlinear systems with non-triangular structural uncertainties,''
  \emph{IEEE Transactions on Automatic Control}, vol.~62, no.~10, pp.
  5220--5226, 2016.

\bibitem{rios2017time}
H.~R{\'\i}os, D.~Efimov, J.~A. Moreno, W.~Perruquetti, and J.~G.
  Rueda-Escobedo, ``Time-varying parameter identification algorithms: Finite
  and fixed-time convergence,'' \emph{IEEE Transactions on Automatic Control},
  vol.~62, no.~7, pp. 3671--3678, 2017.

\bibitem{chen2020adaptive}
K.~Chen and A.~Astolfi, ``Adaptive control for systems with time-varying
  parameters,'' \emph{IEEE Transactions on Automatic Control}, vol.~66, no.~5,
  pp. 1986--2001, 2020.

\bibitem{YAO2021527}
Y.~Yao, J.~Tan, J.~Wu, and X.~Zhang, ``Event-triggered fixed-time adaptive
  neural dynamic surface control for stochastic non-triangular structure
  nonlinear systems,'' \emph{Information Sciences}, vol. 569, pp. 527--543,
  2021.

\bibitem{yao2021event}
------, ``Event-triggered fixed-time adaptive neural tracking control for
  stochastic non-triangular structure nonlinear systems,'' \emph{Neural
  Computing and Applications}, vol.~33, no.~22, pp. 15\,887--15\,899, 2021.

\end{thebibliography}
\end{document}